%% file: main.tex
\documentclass[aps,prb,onecolumn,notitlepage,11pt,longbibliography]{revtex4-2}
\usepackage[T1]{fontenc}
\usepackage{amsmath,amssymb,bm,graphicx,booktabs,mathtools}
\usepackage[colorlinks=true,linkcolor=blue,citecolor=blue,urlcolor=blue]{hyperref}
\newcommand{\hh}{\mathfrak h}
\newcommand{\dd}{\mathrm d}
\newcommand{\id}{I}
\newcommand{\LLL}{\mathrm{LLL}}
\newcommand{\ket}[1]{\lvert #1\rangle}
\newcommand{\bra}[1]{\langle #1\rvert}
\newcommand{\gap}{\operatorname{gap}}

\newcommand{\SM}{Supplemental Material}
\newcommand{\Fock}{\mathcal F}
\newcommand{\Ran}{\operatorname{Ran}}
\newcommand{\Ker}{\operatorname{ker}}
\newcommand{\tr}{\operatorname{tr}}

\newcommand{\Ltwo}{\mathcal L_2}
\begin{document}
\let\OriginalAuthor\author
\let\OriginalAffiliation\affiliation
\let\OriginalMaketitle\maketitle
\let\OriginalAnd\and
\input{article.tex}
\clearpage
\newpage
\let\author\OriginalAuthor
\let\affiliation\OriginalAffiliation
\let\maketitle\OriginalMaketitle
\let\and\OriginalAnd
\makeatletter
\frontmatter@init
\let\@arstrut\@arstrut@org
\let\@arstrut@hook\@empty
\global\setbox\absbox\box\voidb@x
\let\OriginalLabel\label
\expandafter\let\csname package@label@FirstPage\endcsname\relax
\expandafter\let\csname package@label@LastBibItem\endcsname\relax
\renewcommand{\label}[1]{\ifcsname package@label@#1\endcsname\OriginalLabel{supp:#1}\else\OriginalLabel{#1}\fi}
\makeatother
\setcounter{section}{0}
\setcounter{subsection}{0}
\setcounter{subsubsection}{0}
\setcounter{equation}{0}
\setcounter{figure}{0}
\setcounter{table}{0}
\setcounter{footnote}{0}
\setcounter{page}{1}
\renewcommand{\baselinestretch}{1.05}
\renewcommand{\thesection}{S\arabic{section}}
\renewcommand{\thesubsection}{\thesection.\arabic{subsection}}
\renewcommand{\theequation}{S\arabic{equation}}
\renewcommand{\thetable}{S\arabic{table}}
\renewcommand{\thepage}{S\arabic{page}}
\renewcommand{\theHsection}{supp.\arabic{section}}
\renewcommand{\theHsubsection}{\theHsection.\arabic{subsection}}
\renewcommand{\theHsubsubsection}{\theHsubsection.\arabic{subsubsection}}
\renewcommand{\theHequation}{supp.\arabic{equation}}
\renewcommand{\theHfigure}{supp.\arabic{figure}}
\renewcommand{\theHtable}{supp.\arabic{table}}
\normalsize
\input{supplement/supplement.tex}
\makeatletter
\let\auto@bib\relax
\makeatother
\end{document}

%% file: article.tex
\title{Evaluate Laughlin spectral gap lower bound in vortexable bands}
\author{Yuchen Liu}
\affiliation{Kavli Institute for Theoretical Sciences and School of Quantum, University of Chinese Academy of Sciences, Beijing 100190, China}
\author{Zheng Zhu}
\affiliation{Kavli Institute for Theoretical Sciences and School of Quantum, University of Chinese Academy of Sciences, Beijing 100190, China}
\date{\today}
\begin{abstract}

Vortexable Chern bands admit exact Laughlin-type zero modes of local parent interactions, but their existence alone does not guarantee a finite excitation gap. Here, we derive a computable, particle-number-independent lower bound on the spectral gap of Laughlin parent Hamiltonians in a class of vortexable bands obtained by multiplying lowest-Landau-level wavefunctions by a spatially varying envelope. The bound depends only on two quantities: the spread of projected orbital normalizations and a lower bound on the weighted local pair repulsion, both determined by one- and two-particle operators. The proof employs an auxiliary fermion-loss construction to control orbital orthogonalization, with an error bounded relative to the interaction energy uniformly in particle number. We demonstrate that Landau-level averaging permits strictly positive gap bounds even under substantial spatial modulation, including envelopes with nodal lines. 
Beyond the scalar-envelope construction, we extend the analysis to multicomponent orbital textures and deformations of the vortex coordinate.These results establish rigorous gap criteria for Laughlin parent Hamiltonians, providing a foundation for understanding the stability of fractional Chern insulators in vortexable bands.

\end{abstract}
\maketitle
\input{sections/introduction}

\input{sections/criterion}

\input{sections/channel}
\input{sections/examples}

\input{sections/discussion}

\begin{acknowledgments}
Generative AI tools provided assistance with analytical derivations, the development of verification code, and manuscript preparation. The authors have reviewed and verified the resulting calculations and take full responsibility for the scientific content of this work.
\end{acknowledgments}
\section*{Data and code availability}
Supporting derivations are given in the accompanying \SM. Reproducible checks and exact numerical certificates are available at \url{https://github.com/Yuchenian/vortexable-laughlin-gap}.
\input{article.bbl}

%% file: sections/introduction.tex
\section{Introduction}
\label{sec:intro}

The fractional quantum Hall effect (FQHE) demonstrates how interactions can produce incompressible quantum liquids with fractionalized excitations~\cite{Tsui1982,Laughlin1983}. The Laughlin wave function describes their ground-state correlations, while quasihole constructions reveal fractional charge and anyonic statistics~\cite{Laughlin1983,Arovas1984}. Haldane's pseudopotentials and short-range parent interactions make the fermionic $\nu=1/3$ Laughlin state an exact zero-energy ground state of the repulsive $V_1$ Hamiltonian~\cite{Haldane1983,Trugman1985}. Exact ground-state information, however, does not establish a nonzero excitation gap in the thermodynamic limit. Such a gap excludes arbitrarily low-energy excitations above the ground state and is a prerequisite for establishing stability against sufficiently weak local perturbations.

Fractional Chern insulators (FCIs) extend FQH physics to partially filled topological bands without requiring an external magnetic field~\cite{Tang2011,Sun2011,Neupert2011,Sheng2011,Regnault2011}. Quantum-geometric criteria and exact Landau-level representations identify special bands in which this correspondence can be made particularly explicit~\cite{Roy2014,Wang2021,Wang2023,CanoWang2026}. Vortexability provides a constructive real-space perspective~\cite{Ledwith2023}. In the planar lowest Landau level (LLL), multiplying a wave function by a complex coordinate $z-z_0$ inserts a vortex while preserving the Landau-level constraint. A vortexable band admits an analogous operation for a suitable complex coordinate $\zeta(\bm r)$, allowing Laughlin-type correlation zeros to be introduced without leaving the band. At appropriate fillings and in the absence of band dispersion, these wave functions are exact zero modes of suitable positive-semidefinite parent interactions obtained by projecting local interactions into the band~\cite{Ledwith2020,Ledwith2023,Wang2023}.

For suitable ideal Chern bands, the correspondence with the LLL can also be formulated using a nonuniform magnetic field and a spatially varying metric~\cite{Estienne2023}. Such structures arise in controlled limits of models relevant to moir\'e materials. Twisted bilayer graphene in the chiral limit has exactly flat, vortexable bands at magic angles~\cite{Tarnopolsky2019,Ledwith2020}. Aharonov--Casher zero modes in a nonuniform magnetic field~\cite{Aharonov1979} likewise provide ideal reference bands for adiabatic descriptions of twisted transition-metal dichalcogenide homobilayers~\cite{MoralesDuran2024,Shi2024}. These models are relevant to the fractional quantum anomalous Hall states observed in twisted MoTe$_2$~\cite{Cai2023,Zeng2023,Park2023,Xu2023}. Although these ideal limits do not by themselves describe all material effects, they offer analytically tractable starting points for understanding the correlated phases.

Nevertheless, exact Laughlin-type zero modes do not by themselves guarantee a finite thermodynamic excitation gap. Even in ideal Chern bands, the momentum occupations of Laughlin-type states can vary with band geometry~\cite{Liu2025}. For Coulomb interactions, numerical studies show that increasing Berry-curvature inhomogeneity can suppress the neutral gap and favor charge-density-wave order despite the persistence of ideal quantum geometry~\cite{Shi2026}. More directly at the wave-function level, plasma-mapping analyses identify Laughlin-type states in strongly inhomogeneous ideal bands with algebraically decaying correlations and gapless behavior~\cite{Moitra2026}. Although these results concern different interactions and diagnostics, they highlight the need to distinguish exact zero-mode constructions from spectral-gap estimates for a specified parent Hamiltonian.

Recent rigorous advances in the Landau-level problem provide a starting point for addressing this question. A Fock-space spectral inequality established a system-size-independent gap for the fermionic Laughlin $V_1$ Hamiltonian on sufficiently large spheres~\cite{OpenAI2026}, while a complementary semidefinite-programming approach developed a hierarchy of rigorous gap bounds~\cite{Scheer2026}. Subsequent improvements yielded a certified lower bound of $\Delta_0=1/5$ on sufficiently large spheres, together with a uniform positive spectral bound for the untruncated planar Hamiltonian~\cite{Gao2026}. These advances provide a rigorous reference gap and motivate the question of whether such a bound can be extended to vortexable deformations of the lowest Landau level.

In this work, we derive such a bound for local Laughlin parent Hamiltonians in the class of vortexable one-particle spaces $\hh_f=f\hh_{\LLL}$, obtained by multiplying LLL wave functions by a scalar spatial envelope. Starting from a certified reference gap, the bound depends only on the spread of projected orbital normalizations and a lower bound on the weighted local pair penalty. Evaluating these one- and two-particle quantities requires neither a new many-body diagonalization nor a finite-size extrapolation. Whenever the resulting bound is positive, it certifies a particle-number-independent gap above the full zero-mode manifold on the complete plane. The resulting sufficient criterion accommodates examples with strong short-wavelength modulation and even nodal envelopes: LLL averaging can keep the projected normalization close to a scalar while retaining a strictly positive weighted pair penalty. The proof compares orbital orthogonalization with an auxiliary fermion-loss process, with an error bounded by the interaction energy uniformly in particle number. We illustrate these possibilities and discuss extensions to orbital textures and changes of the vortex coordinate. This provides an explicit energetic criterion for gapped Laughlin parents, without assuming that vortexability alone guarantees a stable FCI phase.

%% file: sections/criterion.tex
\section{Model and main result}
\label{sec:setting}
\subsection{Band and parent Hamiltonian}
We consider spin-polarized fermions in a flat band obtained by multiplying planar LLL wave functions by a spatially varying envelope,
\begin{equation}
 \psi(\bm r)=f(\bm r)\phi(\bm r),\qquad \phi\in\hh_{\LLL}.
 \label{eq:band}
\end{equation}
Here $f$ is a smooth, bounded scalar function. Since $f$ commutes with the complex coordinate $z=x+iy$, vortex attachment still acts within this band. More generally, a band projector $P$ is vortexable when $(1-P)\zeta P=0$ for an appropriate complex coordinate $\zeta(\bm r)$~\cite{Ledwith2023}. The deformation in Eq.~\eqref{eq:band} retains $\zeta=z$. 

Such a construction arises naturally in the Aharonov–Casher(AC) problem~\cite{Aharonov1979}.  
For one spin branch of the Pauli Hamiltonian with $g=2$,
\begin{equation}
 H_{\rm AC}=\frac{\bm\pi^2-B(\bm r)}{2m},\qquad
 \bm\pi=-i\bm\nabla-\bm A,
 \label{eq:AC}
\end{equation}
in units $\hbar=q=1$~\cite{Aharonov1979}, a field $B=B_0+\nabla^2\widetilde\varphi$ with periodic $\widetilde\varphi$ yields zero modes of the form~\eqref{eq:band}, with $f=e^{-\widetilde\varphi}$ and magnetic length $\ell=B_0^{-1/2}$. The scalar envelope $f$ therefore captures a physically motivated class of nonuniform-field bands without altering their holomorphic structure.

We study the repulsive derivative contact interaction projected into this band,
\begin{equation}
 H_f=4\pi\ell^4\int\dd^2r\sum_{j=x,y}
 [\Psi_f\partial_j\Psi_f]^\dagger[\Psi_f\partial_j\Psi_f],
 \label{eq:contact}
\end{equation}
where $\Psi_f(\bm r)$ is the band-projected fermion field. For $f=1$, our normalization gives the LLL $V_1$ parent Hamiltonian $H_0=\sum_{i<j}P_{ij}^{(1)}$, with unit energy for a pair of relative angular momentum one~\cite{Haldane1983,Trugman1985}.
Physically, the contact interaction penalizes the term linear in the relative coordinate of two fermions. Since multiplication by $f$ leaves its third-order zeros at particle coincidence intact, the deformed Laughlin state 
\begin{equation}
 \Psi _{L,f}(\boldsymbol{r}_1,...,\boldsymbol{r}_N)\propto \left[ \prod_{i<j}{\left( z_i-z_j \right) ^3} \right] \prod_i{f(\boldsymbol{r}_i)\,e^{-\left| z_i \right|^2/4\ell ^2}},
 \label{eq:Laughlin}
\end{equation}
remains an exact zero mode of $H_f$. This construction establishes the zero-energy states, but does not by itself  determine the gap above the zero-mode manifold.

\subsection{Gap lower bound}
The excitation energy of Eq.~\eqref{eq:contact} depends on more than these zeros. The estimation of the excitation gap depends on two effects: the modification of single-particle normalization and the weighting of the local pair repulsion. 

Multiplying all orbitals by $f$ changes their inner products. The operator
\begin{equation}
 G=P_0|f|^2P_0,\qquad 0< g_-\id\preceq G\preceq g_+\id,
 \label{eq:gram}
\end{equation}
measures this change within the reference LLL, with $P_0$ the LLL projector. $g_{-}$ and $g_{+}$ are positive lower and upper spectral bounds on $G$, respectively. We define $\delta=1-g_-/g_+$ to quantify the nonuniformity of orbital
normalization. A small $\delta$ means all LLL orbitals acquire nearly the same normalization.  
The properly orthonormalized map from LLL to deformed LLL is $U=fG^{-1/2}$. It is unitary between the two one-particle spaces, even though multiplication by $f$ alone is not.

Besides the orbital normalization, the band deformation changes the effective strength of the local pair penalty which gives an overall scaling of the spectral gap. 
Writing the projected field as $\Psi_f=f\chi$, where $\chi$ includes the orthogonalization factor $G^{-1/2}$, we have $\chi(\mathbf r)^2=0$ and hence $(f\chi)\partial_j(f\chi)=f^2\chi\,\partial_j\chi$.
The derivative of the envelope therefore does not generate an
additional contact term. The pair interaction instead acquires
the local weight $|f(\mathbf r)|^4$.  
To quantify this weight, resolve the $V_1$ pair projector into localized relative-angular-momentum-one pairs $p_{\bm R}$, centered at $\bm R$, with measure $\dd\nu=\dd^2R/(\pi\ell^2)$. For LLL and deformed LLL, the corresponding local pair penalty are measured by
\begin{equation}
 Q_1=\int\dd\nu\,
       \ket{p_{\bm R}}\bra{p_{\bm R}}
       \label{eq:Q1}
\end{equation}
and
\begin{equation}
 Q_f=\int\dd\nu\,|f(\bm R)|^4
       \ket{p_{\bm R}}\bra{p_{\bm R}},
 \label{eq:Qf}
\end{equation}
respectively. We search for a two-side bound of $Q_f$:
\begin{equation}
 \kappa Q_1\preceq Q_f\preceq KQ_1.
 \label{eq:pairweight}
\end{equation}
For bounded $f$, the upper bound holds with $K=\|f\|_\infty^4$, while $\kappa>0$ is the nontrivial condition. The constant $\kappa$ gives a uniform lower bound on the weighted contact energy of normalized states in the $V_1$ pair subspace. {A simple sufficient condition is $m=\inf_{\bm R}|f(\bm R)|^4>0$, in which case one may take $\kappa=m$. This pointwise condition is not necessary. The contact energy averages $|f|^4$ over the spatial distribution of the coherent-pair amplitudes, so a positive $\kappa$ may exist even when $f$ has zeros.}

Under these conditions, we obtain the spectral-gap bound
\begin{equation}
 \Delta_f\geq\frac{\kappa}{g_-^2}
          \left(\Delta_0-8\sqrt{3}\delta-\delta^2\right),
 \label{eq:gapbound}
\end{equation}
where $\Delta_0$ can be any certified spectral gap lower bound for the fermionic Laughlin $V_1$ Hamiltonian. Whenever the right-hand side is positive, it gives a particle-number-independent gap $\Delta_f$ above the full zero-mode space of Eq.~\eqref{eq:contact}. This zero space is obtained by multiplying the reference zero modes by $\prod_i f(\bm r_i)$, as in Eq.~\eqref{eq:Laughlin}.  
Equation~\eqref{eq:gapbound} separates the effects of orbital normalization inhomogeneity, controlled by $\delta$, from renormalized contact penalty, characterized by $\kappa/g_-^2$. In particular, a positive parent-gap bound can persist even when the envelope vanishes along nodal lines, as explicitly demonstrated in Sec.~IV.

%% file: sections/channel.tex
\section{Overview of the proof}
\label{sec:mechanism}
A direct comparison of many-body wavefunction norms is insufficient to obtain a particle-number-independent gap bound. Indeed, the corresponding estimates can scale as $(g_+/g_-)^{N/2}$. We instead exploit the two-body structure of the parent Hamiltonian to control the effect of orbital orthogonalization directly at the level of interaction energy.

\subsection{Fermion-loss representation of orbital orthogonalization}
To control the orthogonalization factor $G^{-1/2}$, we introduce the positive contraction
\begin{equation}
 M=\sqrt{g_-}\,G^{-1/2},\qquad 0\prec M \preceq \id.
 \label{eq:M}
\end{equation}
Add an auxiliary copy of $\hh_{\LLL}$. Let $c_{\rm sys}(v)$ and $d_{\rm env}(v)$ denote the system and auxiliary annihilation operators, with fermionic anticommutation relations also between the two subsystems. For $v\in\hh_{\LLL}$, set
\begin{align}
 E&=(\id-M^2)^{1/2},\nonumber\\
 b(v)&=c_{\rm sys}(Mv)+d_{\rm env}(Ev).
 \label{eq:lossmodes}
\end{align}
Write $A(b)$ for the operator obtained from $A$ by replacing every reference mode $c(v)$ and its adjoint by $b(v)$ and its adjoint. We define
\begin{equation}
 \Phi_M(A)={}_{\rm env}\!\bra{0}A(b)\ket{0}_{\rm env}.
 \label{eq:Phi_M}
\end{equation}
Since $M^2+E^2=I$, the operators $b(v)$ satisfy the canonical anticommutation relations. Taking the auxiliary vacuum expectation therefore defines a unital completely positive map $\Phi_M$, corresponding to fermionic particle loss in the Heisenberg picture.

The reference gap can be expressed as $H_0^2\succeq\Delta_0H_0$: for every eigenvalue $\mathcal{E}$, it excludes the interval $0<\mathcal{E}<\Delta_0$. Vacuum averaging preserves this inequality, so
\begin{equation}
 \Phi_M(H_0^2)\succeq\Delta_0H_M,\, H_M=\Phi _M\left( H_0 \right) 
 \label{eq:mappedgap}
\end{equation}
However, an average of a square is not the square of the average. Their difference is a positive fluctuation term,
\begin{equation}
 \Phi_M(H_0^2)=H_M^2+\mathcal N_M.
 \label{eq:variance}
\end{equation}
If it is at most $\eta H_M$, Eqs.~\eqref{eq:mappedgap} and \eqref{eq:variance} immediately give a gap at least $\Delta_0-\eta$ for $H_M$.

\subsection{An interaction-relative bound on the channel variance}
Acting on the auxiliary vacuum, the dilated two-body Hamiltonian annihilates a pair of system fermions and recreates it in one of three mutually orthogonal sectors, containing zero, one, or two auxiliary fermions. The zero-auxiliary sector contributes $H_M^2$, while the remaining sectors give
\begin{equation}
 \mathcal N_M=X_1^\dagger X_1+X_2^\dagger X_2.
 \label{eq:noises}
\end{equation}

Both error terms contain the same pair-annihilation
map that defines $H_M$. Their auxiliary factors
satisfy
\begin{equation}
 \|E\|^2=\|\id-M^2\|\leq\delta,
 \qquad
 \|\wedge^2 E\|^2\leq\delta^2.
 \label{eq:norm_of_loss}
\end{equation}
The remaining factors involve fermionic creation
operators, whose norms are independent of particle
number. The two-auxiliary contribution is controlled
by the pair-frame resolution, while the one-auxiliary
contribution is bounded using a Schur estimate for
overlaps between localized two-frames. Their
Gaussian decay on the magnetic-length scale yields
a finite, system-size-independent coefficient.

The resulting operator bound is
\begin{equation}
 \mathcal N_M\preceq(8\sqrt{3}\delta+\delta^2)H_M.
 \label{eq:relative}
\end{equation}
The constant $8\sqrt{3}$ is a convenient bound on the integrated overlap of coherent $V_1$ pairs. Its derivation and the explicit amplitudes $X_1,X_2$ are given in the \SM. In particular, the error vanishes on zero modes, where no penalized pair can be removed.

We finally relate $H_M$ to the physical contact Hamiltonian $H_f$. In the orthonormal reference representation, Eq.~\eqref{eq:pairweight} gives
\begin{equation}
 \frac{\kappa}{g_-^2}H_M\preceq U_N^\dagger H_fU_N\preceq \frac{K}{g_-^2}H_M,
 \qquad U_N=\bigwedge^N U.
 \label{eq:comparison}
\end{equation}
The two-sided comparison establishes equality of the kernels. Combining the lower bound with the spectral inequality for $H_M$ proves Eq.~\eqref{eq:gapbound}, uniformly in particle number.

%% file: sections/examples.tex
\section{Examples and extensions}
\label{sec:examples}

The gap criterion in Eq.~\eqref{eq:gapbound} depends on projected one- and two-particle quantities, rather than directly on the pointwise amplitude of the spatial deformation. Periodic and nodal envelopes illustrate how Landau-level averaging yields explicit estimates of these one- and two-particle quantities. We then extend the envelope picture to orbital textures and deformed vortex coordinates. Throughout this section, we use the certified reference bound $\Delta_0=1/5$~\cite{Gao2026}; the
results are summarized in Table~\ref{tab:bounds}.

\subsection{Landau-level averaging of a periodic modulation}
Periodic $f$ is a natural setup for Chern band, and the distinction between a real-space envelope and its projected normalization is especially transparent for a periodic $f$. A Fourier component with wavevector $\bm q$ enters the LLL Gram operator with the familiar form factor $e^{-q^2\ell^2/4}$. For an envelope normalized so that the spatial mean of $|f|^2$ is one,
\begin{equation}
 \|G-\id\|\leq\alpha,
 \qquad
 \alpha=\sum_{\bm q\ne0}\bigl|(|f|^2)_{\bm q}\bigr|
                         e^{-q^2\ell^2/4}.
 \label{eq:alpha}
\end{equation}
Thus, short-wavelength variations in $|f|^2$ have only a weak effect on the projected Gram operator. We can take $g_\pm=1\pm\alpha$ and $\delta=2\alpha/(1+\alpha)$.

For example, choose the positive square root of
\begin{align}
 |f_{\epsilon,n}(x,y)|^2
 &=1+\frac\epsilon2[\cos(q_nx)+\cos(q_ny)],\nonumber\\
 q_n&=\frac{n\sqrt{2\pi}}{\ell}.
 \label{eq:periodic}
\end{align}
The projected change satisfies $\alpha\leq\epsilon e^{-n^2\pi/2}$, while $\kappa=(1-\epsilon)^2$ follows immediately from the minimum of $|f|^4$. At $n=2$ and $\epsilon=1/2$, $|f|^2$ varies between $1/2$ and $3/2$, whereas $\|G-I\|<10^{-3}$. Equation~\eqref{eq:gapbound} then gives
\begin{equation}
 \Delta_f\geq0.0436.
 \label{eq:positivegap}
\end{equation}
Moreover, the same lower bound remains valid while $\epsilon$ is increased continuously from zero to $1/2$ at fixed $n=2$. This provides a continuous family of gapped parent Hamiltonians starting from the uniform LLL, excluding possible phase transition.

The modulation is compatible with a square magnetic cell of area $2\pi\ell^2$, containing one flux quantum. This gives a rank-one magnetic Bloch band, with Laughlin filling corresponding to one electron per three cells. Higher harmonics change the envelope inside that cell without changing this filling interpretation.

\begin{table}[t]
\caption{Examples of positive parent-gap lower bounds in units where a reference $V_1$ pair projector has energy one. Decimal entries are rounded downward. The planar results are uniform in particle number.}
\label{tab:bounds}
\begin{ruledtabular}
\begin{tabular}{lc}
Deformation & Lower bound\\
\hline
First harmonic: $\epsilon=0.01$, $n=1$ & $0.1402$\\
Second harmonic: $\epsilon=0.5$, $n=2$ & $0.0436$\\
Nodal envelope, Eq.~\eqref{eq:nodal} & $0.1786$\\
Mass anisotropy & $w_{F_\lambda}/5$
\end{tabular}
\end{ruledtabular}
\end{table}

\subsection{A band envelope with nodal lines}
Consider the smooth, sign-changing envelope
\begin{equation}
 f(x,y)=2\cos(kx)\cos(ky),\qquad
 k=\frac{\sqrt{2\pi}}{\ell}.
 \label{eq:nodal}
\end{equation}
The envelope now vanishes along a grid of lines. An estimate based on the minimum of $|f|$ would give zero, although an LLL pair cannot be concentrated on those lines with arbitrary precision.

Both quantities in the gap criterion reflect this spatial averaging.
The nonconstant harmonics of $|f|^2$ start at wavevector $2k$,
giving $\|G-\id\|<0.004004$.
To estimate the pair penalty, note that the relative angular
momentum is fixed to one, while the pair center of mass forms
an LLL with magnetic length $\ell/\sqrt2$.
Writing
$w=|f|^4=\sum_{\bm q}w_{\bm q}e^{i\bm q\cdot\bm R}$,
each Fourier component of the pair weight is therefore
attenuated by $e^{-q^2\ell^2/8}$.
The remaining guiding-center translations are unitary,
so the triangle inequality gives
\begin{equation}
 Q_f\succeq\kappa_{\rm F}Q_1,\qquad
 \kappa_{\rm F}
 =w_{\bm0}-\sum_{\bm q\ne0}|w_{\bm q}|
                           e^{-q^2\ell^2/8}.
 \label{eq:nodal_fourier_bound}
\end{equation}
For the envelope in Eq.~\eqref{eq:nodal}, this estimation gives $\kappa>1.98156$. Together these estimates yield
\begin{equation}
 \Delta_f\geq0.1786.
 \label{eq:nodalbound}
\end{equation}
The conservative numerical inputs and their exact-arithmetic verification are given in the \SM.

Despite its nodal lines, the envelope produces only a small change in the projected Gram operator, while the weighted pair interaction retains a positive lower bound. Thus, pointwise zeros of the envelope do not preclude a uniform parent gap.

\subsection{Orbital texture and the vortex coordinate}
\label{sec:geometry}
Vortexable bands need not have the scalar-envelope form of Eq.~\eqref{eq:band}. We consider two physically motivated extensions: an internal orbital texture, as in chiral twisted graphene, and a change of vortex coordinate, as in the curved-space Landau-level description of ideal Chern bands. Although the details differ, both can be understood through the envelope picture.

In chiral twisted bilayer graphene (CTBG), a sublattice-polarized magic-angle band in one spin and valley sector has wavefunctions
\begin{equation}
 \boldsymbol\psi_{\bm k}(\bm r)
 =\mathcal N_{\bm k}\,t(\bm r)\Phi_{\bm k}^{\rm LLL}(\bm r),
 \label{eq:ctbgspinor}
\end{equation}
where the two-component layer spinor $t$ is independent of momentum~\cite{WangZheng2021}. Thus the LLL correspondence survives even when the electron's layer composition varies across a moir\'e cell. For the orbital-symmetric derivative contact, summed over layer indices, writing $\Psi_a=t_a\chi$ gives
$\Psi_a\partial_j\Psi_b=t_at_b\chi\partial_j\chi$ due to derivatives of $t$ multiply $\chi^2=0$. The criterion consequently depends on $w=t^\dagger t$ through the Gram operator $P_0wP_0$ and the pair weight $w^2$. Redistributing amplitude between layers at fixed $w$ leaves this parent spectrum unchanged. The \SM\ develops this connection and a generalized chiral Dirac model in which the total weight can be tuned towards a positive gap bound.

A second extension concerns the vortex coordinate itself. This is natural in descriptions of ideal Chern bands as Landau levels in a nonuniform geometry~\cite{Estienne2023}. A smooth, globally invertible, orientation-preserving coordinate transformation $\bm R=F(\bm r)$ carries the LLL into a new subspace by
\begin{equation}
 (U_F\phi)(\bm r)=\sqrt{\det DF(\bm r)}\,\phi(F(\bm r)).
 \label{eq:pullback}
\end{equation}
Since $U_F$ preserves orbital inner products, no additional orthogonalization is needed.With the physical contact~\eqref{eq:contact} held fixed,
its expression in the reference LLL becomes
\begin{align}
 U_{F,N}^\dagger H_FU_{F,N}
 &=\int\dd\nu(\bm R)\,
       w_F(\bm R)B_{\bm R}^\dagger B_{\bm R},
 \label{eq:coordinate_pair}\\
 w_F(\bm R)
 &=\left.
   \frac{J(\bm r)}{2}\|DF(\bm r)\|_{\rm F}^2
   \right|_{\bm r=F^{-1}(\bm R)}.
 \label{eq:metricweight}
\end{align}
Here $U_{F,N}=\wedge^NU_F$ and
$B_{\bm R}$ removes a reference
$V_1$ pair $p_{\bm R}$ centered at $\bm R$.
Thus the coordinate deformation only changes the local strength
of the pair repulsion.
The weight follows from the Jacobian and the transformation
of the derivatives in the contact; its derivation is given
in the \SM.

Mass anisotropy provides a simple physical example of this general construction, closely related to the metric description of FQH states~\cite{Haldane2011,Qiu2012}. Consider electrons in a uniform perpendicular magnetic field with
\begin{equation}
 h=\frac{\pi_x^2}{2m_x}+\frac{\pi_y^2}{2m_y},
 \qquad m_x,m_y>0.
 \label{eq:anisotropic_kinetic}
\end{equation}
Set $\lambda=\tfrac14\ln(m_x/m_y)$.
The area-preserving squeeze $F_\lambda(x,y)=(e^\lambda x,e^{-\lambda}y)$ maps the isotropic LLL of mass $\sqrt{m_x m_y}$ to this anisotropic LLL by
$(U_\lambda\phi)(\bm r)=\phi(F_\lambda(\bm r))$.
The vortex coordinate is now
$\zeta_\lambda=e^\lambda x+i e^{-\lambda}y$.
Keep the isotropic physical contact~\eqref{eq:contact} and its coupling fixed in laboratory coordinates. Its weight in the reference LLL is constant $w_{F_\lambda}=\cosh 2\lambda$. In fact, with $U_{\lambda,N}=\wedge^NU_\lambda$,
\begin{equation}
 U_{\lambda,N}^\dagger H_\lambda U_{\lambda,N}
 =w_{F_\lambda}H_0.
 \label{eq:mass_gap}
\end{equation}
The entire parent spectrum therefore changes its energy scale without a gap closing at any finite positive mass ratio. This is consistent with the complete inheritance of the band-mass anisotropy in the zero-range interaction limit discussed by Krishna \textit{et al.}~\cite{Krishna2019}. The interaction range is essential to this conclusion. Wang \textit{et al.}~\cite{WangAnisotropy2012} found transitions towards density-ordered phases for sufficiently anisotropic Coulomb interactions. Those interactions probe finite pair separations and allow competing correlation geometries.

%% file: sections/discussion.tex
\section{Discussion}
Vortexability allows Laughlin-type wavefunctions to be constructed within a Chern band as exact zero modes of suitable parent interactions. The bound derived here adds a quantitative criterion for retaining a gap as the band changes. For a specified envelope, the task reduces to estimating projected orbital norms and weighted local pair energies. Short-wavelength modulations can be strongly suppressed by Landau-level projection, even when their pointwise amplitudes are large. In particular, an envelope with nodal lines can retain a strictly positive parent gap.

The key to obtaining a uniform gap bound is an interaction-relative estimate on the variance of an auxiliary fermion-loss channel. Fermionic creation-operator bounds and the spatial decay of coherent-pair overlaps ensure that the error remains proportional to the parent interaction, with a coefficient independent of particle number. The same proof strategy extends to interactions admitting suitable localized pair-frame decompositions, as discussed in the Supplemental Material.

The criterion is sufficient but not necessary. A nonpositive lower bound does not establish a gap closing. A concrete next step is to sharpen the estimates of projected orbital norms and the interaction-relative error, retaining more information about the specific band deformation. This could enlarge the regime in which a gap can be certified. The present results concern the specified derivative-contact parents. Studying stability under residual band dispersion and deviations from the parent interaction requires additional analysis, with the certified parent gap providing a quantitative reference.

%% file: supplement/supplement.tex
\title{\texorpdfstring{Supplemental Material for\\Evaluate Laughlin spectral gap lower bound in vortexable bands}{Supplemental Material for Evaluate Laughlin spectral gap lower bound in vortexable bands}}
\author{Yuchen Liu}
\affiliation{Kavli Institute for Theoretical Sciences and School of Quantum, University of Chinese Academy of Sciences, Beijing 100190, China}
\author{Zheng Zhu}
\affiliation{Kavli Institute for Theoretical Sciences and School of Quantum, University of Chinese Academy of Sciences, Beijing 100190, China}
\date{\today}
\maketitle
\raggedbottom
Section~\ref{supp:sec:Criterion} derives the projected contact operator, its lift to many particles, and its zero-mode space. Section~\ref{supp:sec:channel} proves the loss-channel estimate. Sections~\ref{supp:app:norm} and \ref{supp:sec:examples} give the frame normalization and the numerical bounds used in the main text. The remaining sections treat orbital and coordinate deformations, operator domains, and more general pair frames. Equation numbers prefixed by S refer to this supplement. We use the main-text notation $Q_1$, $Q_f$, $\delta$, and $\Delta_0$ throughout the planar application. Machine-checked statements of the gap-transfer argument and its hypotheses are available in the \href{https://github.com/Yuchenian/vortexable-laughlin-gap/tree/main/lean_formalization}{Lean formalization and verification records}.

\section{Reference spectral inequality and scope}
\label{supp:sec:reference}
The external input is the all-sector inequality
\begin{equation}
 H_0=\sum_{i<j}Q_{1,ij},\qquad
 H_0^2\succeq\Delta_0H_0,\qquad \Delta_0=\frac15,
 \label{supp:eq:refgap}
\end{equation}
with coefficient one for each unordered pair. On the plane, $Q_1$ projects onto relative angular momentum one. On a sphere of flux $Q$, the corresponding projector selects pair spin $Q-1$ in the spin-$Q/2$ LLL. The full Fock Hamiltonian is the direct sum of its finite-particle restrictions.

The choice $\Delta_0=1/5$ follows from Theorem~1.1 and Corollary~1.3 of Ref.~\cite{supp:Gao2026}. On the complete plane the inequality holds in every finite-particle sector without an orbital cutoff, above the entire zero-mode space, including quasihole states. On the sphere it holds for all sufficiently large flux, with a threshold independent of particle number. At $Q=3(N-1)$, the spherical zero space is the single Laughlin state. Earlier bounds and related certification methods are given in Refs.~\cite{supp:OpenAI2026,supp:Scheer2026}.

We use this spectral inequality as an input rather than reprove it. In particular, a gap known only at the Laughlin filling would not be enough for the loss-channel step. A periodic envelope on the complete plane does not introduce a finite-torus boundary condition: a torus version would require its own reference inequality and compatible frame estimates. These geometric distinctions are independent of the one-flux magnetic Bloch description of a periodic band.

\input{supplement/sections/S_criterion}
\input{supplement/sections/S_channel}
\input{supplement/sections/S_normalization}
\input{supplement/sections/S_examples}
\input{supplement/sections/S_geometry}

\input{supplement/sections/S_domains}
\input{supplement/sections/S_extensions}
\clearpage
\input{supplement/supplement.bbl}

%% file: supplement/sections/S_criterion.tex
\section{Band projection, the contact interaction, and its zero modes}
\label{supp:sec:Criterion}
\subsection{The band projector and the projected field}
Let $\hh_0=\hh_{\LLL}\subset L^2(\mathbb R^2)$, with orthogonal projector $P_0$ and magnetic length $\ell$. Multiplication by a bounded smooth scalar function $f$ defines $T_f:\hh_0\to L^2$, $T_f\phi=f\phi$. We assume
\begin{equation}
 G=T_f^\dagger T_f=P_0|f|^2P_0\big|_{\hh_0},\qquad
 0<g_-\id\preceq G\preceq g_+\id.
 \label{supp:eq:gram}
\end{equation}
The lower bound makes $T_f$ injective with closed range $\hh_f$. Its polar decomposition gives a unitary map between the two one-particle spaces,
\begin{equation}
 U=T_fG^{-1/2}:\hh_0\longrightarrow\hh_f,\qquad
 U^\dagger U=\id_{\hh_0},\qquad UU^\dagger=P_f.
 \label{supp:eq:U}
\end{equation}
In particular, the orthogonal projector into the deformed band is
\begin{equation}
 P_f=T_fG^{-1}T_f^\dagger
     =fP_0G^{-1}P_0\bar f.
 \label{supp:eq:bandprojector}
\end{equation}
The $P_0$ factors in the last expression specify how $G^{-1}$, an operator on $\hh_0$, is extended to ambient $L^2$. In general $P_f$ is neither $P_0$ nor $fP_0\bar f$.

Choose an orthonormal basis $\{\phi_a\}$ of $\hh_0$ and put $\psi_a=U\phi_a$. Let $a_a$ annihilate the physical orbital $\psi_a$, and let $c_a$ annihilate the reference orbital $\phi_a$. The Fock-space unitary $\Gamma(U)=\bigoplus_N U_N$, $U_N=\wedge^NU$, identifies $a_a$ with $c_a$. Under this identification the projected field becomes
\begin{equation}
 \widetilde\Psi_f(\bm r):=\Gamma(U)^\dagger\Psi_f(\bm r)\Gamma(U)
 =\sum_a\psi_a(\bm r)c_a
 =f(\bm r)\chi(\bm r),\qquad
 \chi(\bm r)=\sum_a(G^{-1/2}\phi_a)(\bm r)c_a.
 \label{supp:eq:projectedfield}
\end{equation}
Thus the band projection is already included in every external orbital. For example,
\begin{equation}
 \{\widetilde\Psi_f(\bm r),\widetilde\Psi_f^\dagger(\bm r')\}
 =P_f(\bm r,\bm r'),\qquad
 \{\chi(\bm r),\chi(\bm r')\}=0.
 \label{supp:eq:fieldCAR}
\end{equation}
The field $\chi$ is not the unfiltered LLL field: its mixed anticommutator is the kernel of $P_0G^{-1}P_0$. Replacing it by the ordinary LLL field would omit orbital orthogonalization.

We define the interaction as the positive projected contact form
\begin{equation}
 H_f=4\pi\ell^4\int\dd^2r\sum_{j=x,y}
 [\Psi_f\partial_j\Psi_f]^\dagger[\Psi_f\partial_j\Psi_f].
 \label{supp:eq:contact}
\end{equation}
Equivalently, the derivative contact is evaluated between wavefunctions whose external legs lie in $\hh_f$. The derivative in this expression differentiates an orbital before evaluating the two particles at coincidence; it is not a prescription to insert another $P_f$ between each field and its derivative. This quadratic-form definition avoids treating a derivative of a delta function as a bounded unprojected potential.

Since $\chi(\bm r)^2=0$, the pullback satisfies
\begin{equation}
 (f\chi)\partial_j(f\chi)=f^2\chi\partial_j\chi,
 \qquad
 \Gamma(U)^\dagger H_f\Gamma(U)
 =4\pi\ell^4\int\dd^2r\,|f|^4
       \sum_j(\chi\partial_j\chi)^\dagger(\chi\partial_j\chi).
 \label{supp:eq:cancel}
\end{equation}
Gauge-connection terms also multiply $\chi^2$ and vanish. Equation~\eqref{supp:eq:cancel} removes derivatives of $f$, but not the dependence of the projection on $f$. The latter remains in $G^{-1/2}$ on all four one-particle legs.

\subsection{From the local contact to a two-particle operator}
Inner products are antilinear in the first argument, so $c(v)$ is antilinear in $v$ and $c^\dagger(v)$ is linear. Exterior products of orthonormal vectors have norm one. For a pair $p=\sum_{a<b}p_{ab}\phi_a\wedge\phi_b$, define
\begin{equation}
 B(p)=\sum_{a<b}\overline{p_{ab}}c_b c_a,
 \qquad B(u\wedge v)=c(v)c(u).
 \label{supp:eq:pairannihilator}
\end{equation}
The pair frame used in the main text is
\begin{equation}
 u_{\bm R}=D(\xi)\ket0,\qquad v_{\bm R}=D(\xi)\ket1,\qquad
 p_{\bm R}=u_{\bm R}\wedge v_{\bm R},\qquad
 \xi=\frac{R_x-iR_y}{\sqrt2\ell},\qquad
 \dd\nu=\frac{\dd^2R}{\pi\ell^2}=\frac{2\dd^2\xi}{\pi}.
 \label{supp:eq:frame}
\end{equation}
Here $D(\xi)=\exp(\xi a^\dagger-\bar\xi a)$ acts on the guiding-center oscillator. The complex conjugation in the definition of $\xi$ makes the frame center equal to the physical point $\bm R$ for the holomorphic orbitals $\phi_n(z)\propto z^n e^{-|z|^2/(4\ell^2)}$ and the antilinear convention for $c(v)$.

Section~\ref{supp:app:norm} proves the resolution
\begin{equation}
 \int\ket{p_{\bm R}}\bra{p_{\bm R}}\dd\nu=Q_1
 \label{supp:eq:Q}
\end{equation}
and the local contact normalization. More explicitly, for a bounded positive one-particle operator $A$ on $\hh_0$, put $\chi_A(\bm r)=\sum_a(A\phi_a)(\bm r)c_a$ and $A_2=\wedge^2A$. In the symmetric gauge,
\begin{equation}
 \chi_A\partial_x\chi_A\big|_{\bm R}
 =-\frac{B(A_2p_{\bm R})}{2\sqrt2\pi\ell^3},\qquad
 \chi_A\partial_y\chi_A=i\chi_A\partial_x\chi_A.
 \label{supp:eq:localpairidentity}
\end{equation}
The identity for general $A$ follows from the unfiltered identity by linearity on the two orbital legs. In particular, $A$ need not commute with position, differentiation, or magnetic translation.

For $A=G^{-1/2}$, Eqs.~\eqref{supp:eq:cancel} and \eqref{supp:eq:localpairidentity} give
\begin{equation}
 \Gamma(U)^\dagger H_f\Gamma(U)
 =\int |f(\bm R)|^4 B(A_2p_{\bm R})^\dagger
                         B(A_2p_{\bm R})\dd\nu.
 \label{supp:eq:pulledcontact}
\end{equation}
Define, as in the main text,
\begin{equation}
 Q_f=\int |f(\bm R)|^4\ket{p_{\bm R}}\bra{p_{\bm R}}\dd\nu,
 \qquad \kappa Q_1\preceq Q_f\preceq KQ_1,
 \qquad 0<\kappa\leq K<\infty.
 \label{supp:eq:coercivity}
\end{equation}
The upper bound holds with $K=\|f\|_\infty^4$. The physical two-particle operator in the orthonormal reference representation is therefore
\begin{equation}
 U_2^\dagger H_f\big|_2 U_2=A_2^\dagger Q_f A_2.
 \label{supp:eq:physicalpair}
\end{equation}
Thus $Q_f$ alone is not the complete pulled-back interaction: the orthogonalization factors on both sides are essential. All operators in Eq.~\eqref{supp:eq:physicalpair} act within $\wedge^2\hh_0$; no additional LLL projection is implicit or omitted.

\subsection{The pair inequality holds in every particle sector}
For a bounded two-particle operator $T$ on $\wedge^2\hh_0$, its pair lift is
\begin{equation}
 \Ltwo(T)\big|_N=\sum_{i<j}T_{ij},\qquad
 \Ltwo(\ket p\bra q)=B(p)^\dagger B(q).
 \label{supp:eq:pairlift}
\end{equation}
The normalization in the second identity follows, for example, from
$B(\phi_a\wedge\phi_b)^\dagger B(\phi_a\wedge\phi_b)=n_an_b$ for $a<b$. 

The map $\Ltwo$ is linear and preserves positivity. Indeed, for any normalized $N$-particle state $\psi$, its unnormalized two-particle reduced density operator $\rho_\psi^{(2)}$ is positive, has trace $\binom N2$, and satisfies
\begin{equation}
 \langle\psi,\Ltwo(T)\psi\rangle
 =\tr\bigl(T\rho_\psi^{(2)}\bigr)\geq0\qquad(T\succeq0).
 \label{supp:eq:liftpositive}
\end{equation}

Set
\begin{equation}
 M=\sqrt{g_-}\,G^{-1/2},\quad M_2=\wedge^2M=g_-A_2,\quad
 H_M=\Ltwo(M_2^\dagger Q_1M_2),\quad
 \delta_M=\|\id-M^2\|\leq\delta:=1-\frac{g_-}{g_+}.
 \label{supp:eq:deltabound}
\end{equation}
First apply the congruence $T\mapsto M_2^\dagger T M_2$ to Eq.~\eqref{supp:eq:coercivity}, and then apply the positive linear map $\Ltwo$. Equations~\eqref{supp:eq:pulledcontact}--\eqref{supp:eq:pairlift} give
\begin{equation}
 \frac{\kappa}{g_-^2}H_M\big|_N
 \preceq U_N^\dagger H_f\big|_N U_N
 \preceq\frac{K}{g_-^2}H_M\big|_N.
 \label{supp:eq:compare}
\end{equation}
The comparison coefficient is unchanged by the lift: both sides are sums over the same unordered pairs. A factor $\binom N2$ enters a bound by the identity, not this relative comparison. The two-body terms need not commute. Taking the direct sum over $N$ yields the corresponding Fock-space form inequality, as detailed in Sec.~\ref{supp:app:closure}.

\subsection{The Laughlin zero and the complete kernel}
For a smooth $N$-particle wavefunction $\Phi$ in $\wedge^N\hh_f$, applying $\Psi_f\partial_j\Psi_f$ evaluates, up to a sign and the factor $\sqrt{N(N-1)}$, the derivative of $\Phi$ in one of its first two coordinates at coincidence. This follows directly by inserting the orthonormal expansion of the projected field. The reproducing projector acts as the identity on each external leg of $\Phi$; there is no approximation in using its coordinate wavefunction.

Write the reference planar Laughlin state as
\begin{equation}
 \Psi_L=\mathcal C\prod_{i<j}(z_i-z_j)^3
             e^{-\sum_i|z_i|^2/(4\ell^2)},\qquad
 \Psi_{L,f}=\mathcal C_f\prod_i f(\bm r_i)\Psi_L.
 \label{supp:eq:Laughlin}
\end{equation}
For fixed remaining coordinates, $\Psi_{L,f}$ is $(z_1-z_2)^3$ times a smooth function of $\bm r_1,\bm r_2$. Hence
\begin{equation}
 \left.\partial_{r_{1,j}}\Psi_{L,f}
 \right|_{\bm r_1=\bm r_2}=0,\qquad j=x,y.
 \label{supp:eq:cubiczero}
\end{equation}
A derivative of the cubic factor leaves a quadratic zero; derivatives of the envelope or the Gaussian leave the cubic factor intact. Every contact annihilator in Eq.~\eqref{supp:eq:contact} therefore kills $\Psi_{L,f}$, and the positive interaction has zero expectation value. Smooth zeros of $f$ do not change this argument. Multiplication by $f$ maps each one-particle leg into $\hh_f$, so this state is in the physical projected Hilbert space, not merely an unprojected trial wavefunction. The same reasoning applies to the usual Laughlin quasihole zero modes.

To identify the entire zero space, rather than only these examples, use both bounds in Eq.~\eqref{supp:eq:compare}. For positive operators they imply
\begin{equation}
 \Ker(U_N^\dagger H_f U_N)=\Ker H_M\big|_N.
 \label{supp:eq:samekernel}
\end{equation}
The lower comparison alone would give only one inclusion. The annihilator identity proved in Sec.~\ref{supp:sec:channel} gives
$\Ker H_M|_N=(\wedge^NM^{-1})\Ker H_0|_N$. Since
$UM^{-1}=T_f/\sqrt{g_-}$, it follows that
\begin{equation}
 \Ker H_f\big|_N
 =\left(\prod_{i=1}^N f(\bm r_i)\right)\Ker H_0\big|_N.
 \label{supp:eq:physicalkernel}
\end{equation}
This step uses the bounded inverse of $M$ on the LLL, not multiplication by $1/f$ on ambient $L^2$.

\subsection{Completing the gap comparison}
The loss-channel estimate established below gives
$H_M^2\succeq(\Delta_0-8\sqrt3\,\delta-\delta^2)H_M$.
When the coefficient is positive, the spectral theorem bounds $H_M$ below by that coefficient on $(\Ker H_M)^\perp$. Equation~\eqref{supp:eq:samekernel} identifies this complement with the physical excited-state subspace under $U_N$. Applying the lower bound in Eq.~\eqref{supp:eq:compare} then yields
\begin{equation}
 \Delta_f\geq b_f:=\frac{\kappa}{g_-^2}
       \bigl(\Delta_0-8\sqrt3\,\delta-\delta^2\bigr),
 \qquad \delta=1-\frac{g_-}{g_+}.
 \label{supp:eq:mainbound}
\end{equation}
For $b_f\leq0$ the criterion supplies no positive gap. The result is uniform in finite particle number and is above the complete zero space, not a selected Laughlin vector on the plane. The maximal domain needed for the statement of vortexability is treated separately in Sec.~\ref{supp:app:domain}.

%% file: supplement/sections/S_channel.tex
\section{The fermion-loss map and the relative variance estimate}
\label{supp:sec:channel}
\subsection{Pair-frame assumptions}
The argument in this section applies to a separable one-particle space $\hh$ and a measurable frame of normalized decomposable pairs $p_x=u_x\wedge v_x$, with $u_x,v_x$ orthonormal. We use the annihilator and lift conventions of Eqs.~\eqref{supp:eq:pairannihilator} and \eqref{supp:eq:pairlift}. Define the synthesis operator and its frame operator by
\begin{equation}
 Sa=\int p_xa(x)\dd\nu(x),\qquad
 Q=SS^\dagger=\int\ket{p_x}\bra{p_x}\dd\nu(x),\qquad
 0\preceq Q\preceq\beta\id.
 \label{supp:eq:generalframe}
\end{equation}
Here $Q$ is a general pair operator; for the planar application $Q=Q_1$ and $\beta=1$. Write $U_x=(u_x,v_x):\mathbb C^2\to\hh$, and suppose a positive weight $\omega$ satisfies
\begin{equation}
 k(x,y)=\|U_x^\dagger U_y\|_1,\qquad
 \int k(x,y)\omega(y)\dd\nu(y)\leq C\omega(x).
 \label{supp:eq:Schur}
\end{equation}
The norm is the trace norm of a two-by-two matrix. Since $k(x,y)=k(y,x)$, this also supplies the second weighted Schur condition. The coherent planar frame has $\omega=1$ and $C\leq8\sqrt3$, as shown in Sec.~\ref{supp:app:norm}.

Let $H_0=\Ltwo(Q)$ satisfy $H_0^2\succeq\Delta_0H_0$ in every finite-particle sector. For a positive contraction $M$, put
\begin{equation}
 E=(\id-M^2)^{1/2},\qquad \delta_M=\|\id-M^2\|,\qquad
 B_x^M=B((\wedge^2M)p_x),\qquad
 H_M=\int(B_x^M)^\dagger B_x^M\dd\nu(x).
 \label{supp:eq:filteredparent}
\end{equation}
We will prove
\begin{equation}
 H_M^2\succeq(\Delta_0-C\delta_M-\beta\delta_M^2)H_M.
 \label{supp:eq:transfer}
\end{equation}
Only the kernel identification at the end requires a bounded inverse for $M$.

\subsection{A direct proof of positivity and complete positivity}
Use the environment-first identification
$\Fock(\hh_{\rm env}\oplus\hh_{\rm sys})
 \simeq\Fock(\hh_{\rm env})\otimes\Fock(\hh_{\rm sys})$.
The operators
\begin{equation}
 d_{\rm env}(v)=d(v)\otimes\id,\qquad
 c_{\rm sys}(v)=(-1)^{N_{\rm env}}\otimes c(v)
 \label{supp:eq:gradedmodes}
\end{equation}
anticommute between the two factors as well as within each factor. Set
\begin{equation}
 Jv=(Ev,Mv),\qquad b(v)=d_{\rm env}(Ev)+c_{\rm sys}(Mv).
 \label{supp:eq:lossmodes}
\end{equation}
Since $E^2+M^2=\id$, $J$ is an isometry and
$\{b(u),b^\dagger(v)\}=\langle u,v\rangle$, $\{b(u),b(v)\}=0$.
Thus replacing $c,c^\dagger$ by $b,b^\dagger$ defines a unital
$*$-representation $\pi_M(A)=A(b)$ of the reference CAR algebra.

This representation also has an explicit unitary realization. The block operator
\begin{equation}
 W=\begin{pmatrix}M&E\\-E&M\end{pmatrix}
 \quad\hbox{on }\hh_{\rm env}\oplus\hh_{\rm sys}
 \label{supp:eq:dilationunitary}
\end{equation}
is unitary because $E$ commutes with $M$. It maps $(0,v)$ to $Jv$, and its fermionic second quantization obeys
$\Gamma(W)c_{\rm sys}(v)\Gamma(W)^\dagger=b(v)$.
This identifies $A(b)$ with the original observable acting on the rotated reference subspace $J\hh$, with its orthogonal complement as a spectator.

Let $V\psi=\ket{0}_{\rm env}\otimes\psi$. The map in the main text is precisely the vacuum compression
\begin{equation}
 \Phi_M(A)=V^\dagger\pi_M(A)V.
 \label{supp:eq:Phi_M}
\end{equation}
For a bounded positive observable $A$, factor $A=T^\dagger T$. Then, for every system vector $\psi$,
\begin{equation}
 \langle\psi,\Phi_M(A)\psi\rangle
 =\|\pi_M(T)V\psi\|^2\geq0.
 \label{supp:eq:positivity}
\end{equation}
This is the required positivity proof. The same argument applied to a positive block matrix $[A_{ij}]$ on an arbitrary finite-dimensional auxiliary space gives
$[\Phi_M(A_{ij})]\succeq0$; hence the map is completely positive.
Also $\Phi_M(\id)=V^\dagger V=\id$. It is therefore a unital completely positive map in the Heisenberg picture; its dual on density matrices is trace preserving. For an eigenmode of $M$ with eigenvalue $\sqrt p$, the unitary mixes the mode with an initially empty auxiliary mode with amplitudes $\sqrt p$ and $\sqrt{1-p}$. This is the particle-loss interpretation.

For the possibly unbounded full Fock Hamiltonian, these manipulations are first made on a fixed total-particle sector. On total particle number $N$, the rotated reference subspace $J\hh$ can contain any $k=0,\ldots,N$ particles. Therefore the reference inequalities in all those sectors imply
\begin{equation}
 H_0(b)^2\succeq\Delta_0H_0(b),\qquad
 \Phi_M(H_0^2)\succeq\Delta_0\Phi_M(H_0).
 \label{supp:eq:mappedgap}
\end{equation}
An inequality known only in the reference $N$-particle sector would not suffice. Section~\ref{supp:app:closure} makes the direct-sum and domain statements explicit.

\subsection{Resolving the channel variance by environment particle number}
Normal ordering and the environment vacuum give
\begin{equation}
 \Phi_M(H_0)=H_M.
 \label{supp:eq:channelmean}
\end{equation}
Indeed, the pair annihilator in $H_0(b)$ first acts on the vacuum and reduces to $B_x^M$; only the two-system-particle creation term survives a subsequent vacuum compression. This does not imply $\Phi_M(H_0^2)=H_M^2$. In the square, an intermediate auxiliary particle can be created and later removed.

Let $\Pi_j$ project onto $j$ environment particles. A normal-ordered two-body Hamiltonian acting on an empty environment has outputs only for $j=0,1,2$. With $X_j=\Pi_jH_0(b)V$ for $j=1,2$,
\begin{equation}
 H_0(b)V=VH_M+X_1+X_2.
 \label{supp:eq:envsplit}
\end{equation}
These three terms have mutually orthogonal ranges. Since $H_0(b)$ is self-adjoint on each fixed sector,
\begin{equation}
 \Phi_M(H_0^2)=H_M^2+\mathcal N_M,\qquad
 \mathcal N_M=X_1^\dagger X_1+X_2^\dagger X_2\succeq0.
 \label{supp:eq:variance}
\end{equation}
In particular, positivity gives the inequality $\Phi_M(H_0^2)\succeq H_M^2$ in the wrong direction to infer a gap directly. The following upper bound on $\mathcal N_M$, relative to $H_M$, is essential.

For fixed $N\geq2$, define the analysis map
\begin{equation}
 A_M:\wedge^N\hh\longrightarrow L^2(X;\wedge^{N-2}\hh),\qquad
 (A_M\psi)(x)=B_x^M\psi,
 \qquad A_M^\dagger A_M=H_M\big|_N.
 \label{supp:eq:analysismap}
\end{equation}
Define also the synthesis map
\begin{equation}
 R_MF=\int\bigl[u_x\otimes c^\dagger(Mv_x)
                  -v_x\otimes c^\dagger(Mu_x)\bigr]F(x)\dd\nu(x).
 \label{supp:eq:RM}
\end{equation}
Its range is $\hh\otimes\wedge^{N-1}\hh$. Expanding the two creation operators in $B_x(b)^\dagger$ gives
\begin{align}
 X_1&=(E\otimes\id)R_MA_M,
       &&X_1:\wedge^N\hh\to\hh_{\rm env}\otimes\wedge^{N-1}\hh_{\rm sys},\nonumber\\
 X_2&=[(\wedge^2E)S\otimes\id]A_M,
       &&X_2:\wedge^N\hh\to\wedge^2\hh_{\rm env}\otimes\wedge^{N-2}\hh_{\rm sys}.
 \label{supp:eq:noises}
\end{align}
The minus sign in Eq.~\eqref{supp:eq:RM} follows from the cross-system anticommutation in Eq.~\eqref{supp:eq:gradedmodes}. Normalized exterior products account for the two-environment amplitude without an additional factor of two.

\subsection{The two-environment bound}
The frame identity $SS^\dagger=Q$ implies $\|S\|^2=\|Q\|\leq\beta$. Also
\begin{equation}
 \|E\|^2=\|\id-M^2\|=\delta_M,
 \qquad \|\wedge^2E\|\leq\|E\|^2=\delta_M.
 \label{supp:eq:lossnorms}
\end{equation}
Consequently,
\begin{equation}
 \|X_2\psi\|^2\leq\beta\delta_M^2\|A_M\psi\|^2
 =\beta\delta_M^2\langle\psi,H_M\psi\rangle,
 \qquad X_2^\dagger X_2\preceq\beta\delta_M^2H_M.
 \label{supp:eq:twoenv}
\end{equation}
The squared loss amplitude is $\delta_M^2$, not $\delta_M$, because both recreated particles enter the environment.

\subsection{The one-environment bound}
For orthonormal two-frames $U=(u,v)$ and $V'=(s,t)$, put
$R_U=u\otimes c^\dagger(Mv)-v\otimes c^\dagger(Mu)$. We first show
\begin{equation}
 \|R_U^\dagger R_{V'}\|\leq\|U^\dagger V'\|_1.
 \label{supp:eq:overlaplemma}
\end{equation}
A change of frame by a two-by-two unitary matrix $L$ gives $R_{UL}=\det(L)R_U$. Independent frame rotations therefore change $R_U^\dagger R_{V'}$ only by a phase. Choose them so that $U^\dagger V'$ is diagonal with singular values $\sigma_1,\sigma_2$. The Gram operator reduces to
\begin{equation}
 \sigma_1c(Mv)c^\dagger(Mt)+\sigma_2c(Mu)c^\dagger(Ms).
\end{equation}
Each product has norm at most one: $M$ is a contraction and the fermionic CAR give $\|c(a)\|=\|a\|$ on the entire Fock space. The triangle inequality yields Eq.~\eqref{supp:eq:overlaplemma}.

For a vector-valued function $F$, the integral Gram kernel of $R_M$ thus obeys
\begin{equation}
 \|R_MF\|^2\leq\iint k(x,y)\|F(x)\|\|F(y)\|
                                   \dd\nu(x)\dd\nu(y).
\end{equation}
Apply $2ab\leq a^2+b^2$ with
$a=\|F(x)\|\sqrt{\omega(y)/\omega(x)}$ and
$b=\|F(y)\|\sqrt{\omega(x)/\omega(y)}$.
Symmetry of $k$ and Eq.~\eqref{supp:eq:Schur} give
$\|R_MF\|^2\leq C\|F\|_{L^2}^2$. Therefore
\begin{equation}
 \|X_1\psi\|^2\leq\|E\|^2\|R_M\|^2\|A_M\psi\|^2
 \leq C\delta_M\langle\psi,H_M\psi\rangle,
 \qquad X_1^\dagger X_1\preceq C\delta_MH_M.
 \label{supp:eq:oneenv}
\end{equation}
Here $E$ was factored out before estimating the overlap kernel. The kernel contains the original local frames $U_x^\dagger U_y$, not matrix elements of $E$ or $M$ between distant frames. No locality or Gaussian decay is assumed for $M=\sqrt{g_-}G^{-1/2}$ itself. The frame overlap supplies the finite constant $C$, and the bounded fermionic creation operators prevent an occupation-number factor.

Combining Eqs.~\eqref{supp:eq:mappedgap}, \eqref{supp:eq:variance}, \eqref{supp:eq:twoenv}, and \eqref{supp:eq:oneenv} proves Eq.~\eqref{supp:eq:transfer}. Every error term is proportional to $H_M$. For the planar application, $\beta=1$, $C\leq8\sqrt3$, and $\delta_M\leq\delta$ give the variance bound used in the main text.

\subsection{Identifying the filtered kernel}
For positive boundedly invertible $M$, contraction of an exterior product gives
\begin{equation}
 B(p)(\wedge^NM)=(\wedge^{N-2}M)B((\wedge^2M)p).
 \label{supp:eq:annihilatorintertwining}
\end{equation}
Because $H_M$ is a sum of squares, $\psi\in\Ker H_M$ exactly when $B_x^M\psi=0$ for almost every $x$. Equation~\eqref{supp:eq:annihilatorintertwining}, together with invertibility of $\wedge^{N-2}M$, makes this equivalent to $(\wedge^NM)\psi\in\Ker H_0$. Hence
\begin{equation}
 \Ker H_M\big|_N=(\wedge^NM^{-1})\Ker H_0\big|_N.
 \label{supp:eq:filterkernel}
\end{equation}
This use of the many-particle inverse identifies a subspace only. It does not estimate its condition number or use it to compare nonzero energies.

%% file: supplement/sections/S_normalization.tex
\section{Coherent-pair normalization and overlap constants}
\label{supp:app:norm}
\subsection{The planar pair resolution}
With $\xi$ as in Eq.~\eqref{supp:eq:frame}, displacement of the oscillator states gives
\begin{equation}
 \langle n,u_{\bm R}\rangle=e^{-|\xi|^2/2}\frac{\xi^n}{\sqrt{n!}},\qquad
 \langle n,v_{\bm R}\rangle=e^{-|\xi|^2/2}
       \frac{n\xi^{n-1}-\bar\xi\xi^n}{\sqrt{n!}}.
\end{equation}
The term with $n=0$ in $n\xi^{n-1}$ is understood as zero. Antisymmetrization cancels the $\bar\xi$ terms, so
\begin{equation}
 \langle i\wedge j,p_{\bm R}\rangle
 =e^{-|\xi|^2}\frac{(j-i)\xi^{i+j-1}}{\sqrt{i!j!}},\qquad i<j.
 \label{supp:eq:paircoefficients}
\end{equation}
Integrating with measure $2\dd^2\xi/\pi$, angular integration separates the blocks $L=i+j$. In each block radial integration gives the rank-one projector onto
\begin{equation}
 q_L(i,L-i)=(L-2i)
 \sqrt{\frac{(L-1)!}{2^{L-1}i!(L-i)!}},\qquad i<L/2.
\end{equation}
The binomial identity
$\sum_{i=0}^L(L-2i)^2\binom Li=L2^L$, together with pairing $i$ and $L-i$, gives $\|q_L\|=1$ for every $L\geq1$.

Equivalently, introduce the center and relative oscillators
$a_{\rm CM}=(a_1+a_2)/\sqrt2$ and $a_{\rm rel}=(a_1-a_2)/\sqrt2$.
At $\bm R=0$, $\ket0\wedge\ket1$ is, up to a constant phase, the state with one relative quantum and no center quantum. Displacing both particles by $\xi$ displaces only the center oscillator by $\sqrt2\xi$. Thus $p_{\bm R}$ has fixed relative angular momentum one and a center coherent state. The integrated blocks exhaust that pair subspace:
\begin{equation}
 \int\ket{p_{\bm R}}\bra{p_{\bm R}}\dd\nu=Q_1,
 \qquad \|Q_1\|=1.
\end{equation}
This establishes both the normalization of the resolution and $\beta=1$ in Eq.~\eqref{supp:eq:generalframe}.

\subsection{Matching the derivative contact}
In the symmetric gauge, use the normalized orbitals
\begin{equation}
 \phi_n(z)=\frac1{\sqrt{2\pi\ell^2n!}}
      \left(\frac z{\sqrt2\ell}\right)^n e^{-|z|^2/(4\ell^2)}.
 \label{supp:eq:orbitals}
\end{equation}
For $\Psi_0=\sum_n\phi_n c_n$, the coefficient of $c_i c_j$ in
$\Psi_0\partial_x\Psi_0$, for $i<j$, is
\begin{equation}
 \phi_i\partial_x\phi_j-\phi_j\partial_x\phi_i
 =\frac{e^{-|z|^2/(2\ell^2)}}{2\sqrt2\pi\ell^3}
       \frac{(j-i)(z/\sqrt2\ell)^{i+j-1}}{\sqrt{i!j!}}.
 \label{supp:eq:contactcoefficients}
\end{equation}
The differentiated Gaussian cancels in this antisymmetric combination. Since $z/\sqrt2\ell=\bar\xi$ at $\bm R$, comparing Eqs.~\eqref{supp:eq:paircoefficients} and \eqref{supp:eq:contactcoefficients} gives
$\Psi_0\partial_x\Psi_0=-(2\sqrt2\pi\ell^3)^{-1}B(p_{\bm R})$.
Holomorphicity gives $\Psi_0\partial_y\Psi_0=i\Psi_0\partial_x\Psi_0$, because the nonholomorphic Gaussian contribution is proportional to $\Psi_0^2$. Hence
\begin{equation}
 4\pi\ell^4\sum_{j=x,y}(\Psi_0\partial_j\Psi_0)^\dagger
                                  (\Psi_0\partial_j\Psi_0)\,\dd^2R
 =B(p_{\bm R})^\dagger B(p_{\bm R})\,\frac{\dd^2R}{\pi\ell^2}.
\end{equation}
This proves the contact coefficient and the measure used in Eq.~\eqref{supp:eq:pulledcontact}. Filtering both orbital legs by a positive operator $A$ replaces $p_{\bm R}$ by $(\wedge^2A)p_{\bm R}$, proving Eq.~\eqref{supp:eq:localpairidentity} as well.

\subsection{The planar Schur constant}
Let $U_\xi=(D(\xi)\ket0,D(\xi)\ket1)$ and
$t=|\xi-\xi'|^2$. Displacement-operator matrix elements give
\begin{equation}
 \|U_\xi^\dagger U_{\xi'}\|_{\mathrm F}^2=e^{-t}(t^2+2),\qquad
 |\det(U_\xi^\dagger U_{\xi'})|=e^{-t}.
\end{equation}
For a two-by-two matrix $A$, $\|A\|_1^2=\|A\|_{\mathrm F}^2+2|\det A|$. Translation invariance and radial integration with $\dd\nu=2\dd^2\xi'/\pi$ therefore give
\begin{align}
 C_{\rm plane}
 &=\int\|U_\xi^\dagger U_{\xi'}\|_1\dd\nu(\xi')
   =2\int_0^\infty e^{-t/2}\sqrt{t^2+4}\,\dd t\nonumber\\
 &\leq2\sqrt{\left(\int_0^\infty e^{-t/2}\dd t\right)
             \left(\int_0^\infty e^{-t/2}(t^2+4)\dd t\right)}
   =8\sqrt3.
 \label{supp:eq:Cplane}
\end{align}
The integrals in the last line are $2$ and $24$. This analytic bound, rather than an approximate quadrature, is the constant used in the main text.

\subsection{The spherical frame and contact normalization}
\label{supp:app:sphere}
In this subsection $Q$ denotes magnetic flux, not the general pair operator of Sec.~\ref{supp:sec:channel}. At flux $Q\geq1$, rotate the highest and next-highest states of spin $Q/2$ together to form $u_g,v_g$. Their wedge is a normalized coherent state of pair spin $Q-1$, whose dimension is $2Q-1$. With normalized Haar measure,
\begin{equation}
 (2Q-1)\int\ket{p_g}\bra{p_g}\dd g=P_{\mathrm{pair\ spin}\ Q-1}.
\end{equation}
Thus $\beta=1$ at every flux. For $s=\sin^2(\theta/2)$, the azimuth-integrated Haar measure is $\dd s$. The two-frame overlap matrix $A$ obeys
\begin{align}
 |\det A|&=(1-s)^{Q-1},\\
 \|A\|_{\mathrm F}^2&=(1-s)^{Q-2}[2-2s+(Q-1)^2s^2].
\end{align}
Consequently,
\begin{align}
 C_Q&=(2Q-1)\int_0^1(1-s)^{(Q-2)/2}
                   \sqrt{4(1-s)+(Q-1)^2s^2}\,\dd s\nonumber\\
 &\leq\frac{2(2Q-1)}Q
       \sqrt{\frac{12Q^2+8}{(Q+2)(Q+4)}}<8\sqrt3.
 \label{supp:eq:CQ}
\end{align}
For the Cauchy--Schwarz step, the weight integrates to $2/Q$, and its normalized moments of $1-s$ and $s^2$ are $Q/(Q+2)$ and $8/[(Q+2)(Q+4)]$. At $Q=1$ the apparent endpoint singularity is removable: the trace norm is constantly two and $C_1=2$.

In local coordinates $\zeta=(X+iY)/(2R)$ about the north pole, the normalized monopole orbitals are
\begin{equation}
 \phi_n=\sqrt{\frac{(Q+1)\binom Qn}{4\pi R^2}}\,
        \frac{\zeta^n}{(1+|\zeta|^2)^{Q/2}}.
\end{equation}
The sum of the squared contact amplitudes in two orthonormal tangent directions is
$Q(Q+1)^2/(32\pi^2R^6)$. Multiplication by $c_Q$ in Eq.~\eqref{supp:eq:spherecontact} gives the coherent-pair density $(2Q-1)/(4\pi R^2)$. With $\ell^2=2R^2/Q$,
$c_Q/(4\pi\ell^4)=Q(2Q-1)/[2(Q+1)^2]\to1$. The finite-flux coefficient, not its planar limit, must be used in the spherical comparison.

%% file: supplement/sections/S_examples.tex
\section{Fourier estimates and the numerical bounds in the main text}
\label{supp:sec:examples}
\subsection{One-particle and pair form factors}
Write the coordinate of a particle as a guiding center plus a cyclotron coordinate. The cyclotron oscillator is in its ground state in the LLL, and its matrix element of $e^{i\bm q\cdot\bm r}$ is $e^{-q^2\ell^2/4}$. Thus, for an absolutely summable Fourier expansion $a(\bm r)=\sum_{\bm q}a_{\bm q}e^{i\bm q\cdot\bm r}$,
\begin{equation}
 P_0aP_0\big|_{\hh_0}=\sum_{\bm q}a_{\bm q}e^{-q^2\ell^2/4}W(\bm q),
 \label{supp:eq:gramFourier}
\end{equation}
where $W(\bm q)$ is a unitary guiding-center translation. When the spatial mean of $|f|^2$ is one, the triangle inequality gives
\begin{equation}
 \|G-\id\|\leq\alpha:=\sum_{\bm q\ne0}
       \bigl|(|f|^2)_{\bm q}\bigr|e^{-q^2\ell^2/4}.
 \label{supp:eq:alpha}
\end{equation}
Any proved upper bound $\alpha\leq a<1$ gives valid inputs
$g_\pm=1\pm a$ and $\delta=2a/(1+a)$. The spectral bounds need not be the optimal endpoints of the spectrum of $G$.

For the pair operator, the relative oscillator is fixed in its angular-momentum-one state. Its remaining degree of freedom is the center coordinate $\bm R=(\bm r_1+\bm r_2)/2$. As shown in Sec.~\ref{supp:app:norm}, its coherent state has oscillator amplitude $\sqrt2\xi$, so the center magnetic length is $\ell_{\rm CM}=\ell/\sqrt2$. Under the identification of $\Ran Q_1$ with this center LLL, the coherent-state resolution gives
\begin{equation}
 \int e^{i\bm q\cdot\bm R}\ket{p_{\bm R}}\bra{p_{\bm R}}\dd\nu
 =e^{-q^2\ell^2/8}W_{\rm CM}(\bm q)
 \quad\hbox{on }\Ran Q_1.
 \label{supp:eq:pairFourier}
\end{equation}
This is the LLL form factor at magnetic length $\ell_{\rm CM}$, not the one-particle form factor at $\ell$. The operator vanishes on $(\Ran Q_1)^\perp$. For $w=|f|^4$,
\begin{equation}
 Q_f\big|_{\Ran Q_1}=\sum_{\bm q}w_{\bm q}
           e^{-q^2\ell^2/8}W_{\rm CM}(\bm q),\qquad
 \kappa=w_{\bm0}-\sum_{\bm q\ne0}|w_{\bm q}|e^{-q^2\ell^2/8}.
 \label{supp:eq:kappaFourier}
\end{equation}
A positive value of this $\kappa$ proves the lower bound in Eq.~\eqref{supp:eq:coercivity}. Replacing the minus sign by a plus sign gives a possible upper bound $K$. No commutativity of the guiding-center translations is required for these norm estimates.

\subsection{Positive periodic envelopes and a common path bound}
For the positive square root of
\begin{equation}
 |f_{\epsilon,n}|^2=1+\frac\epsilon2[\cos(q_nx)+\cos(q_ny)],\qquad
 q_n=\frac{n\sqrt{2\pi}}{\ell},\quad 0\leq\epsilon<1,
 \label{supp:eq:periodic}
\end{equation}
the four nonzero Fourier coefficients have absolute value $\epsilon/4$. Therefore
\begin{equation}
 \alpha=\epsilon e^{-n^2\pi/2},\qquad
 \kappa=(1-\epsilon)^2.
 \label{supp:eq:periodicinputs}
\end{equation}
The second relation is a valid lower bound from the pointwise minimum of $|f|^4$; it need not be the optimal pair bound.

Substituting these inputs into Eq.~\eqref{supp:eq:mainbound} gives the first two bounds in Table~\ref{supp:tab:bounds}. Along the path $n=2$, $0\leq\epsilon\leq1/2$, we can use the common values $a=0.000934$ and $\kappa=1/4$, giving
\begin{equation}
 \Delta_f\geq0.0436
 \label{supp:eq:pathbound}
\end{equation}
throughout. The periodic modulation can therefore be turned on continuously from the LLL while retaining a positive parent gap.

\subsection{The nodal envelope}
For
\begin{equation}
 f(x,y)=2\cos(kx)\cos(ky),\qquad k=\frac{\sqrt{2\pi}}\ell,
 \label{supp:eq:nodal}
\end{equation}
we have $|f|^2=(1+\cos2kx)(1+\cos2ky)$. The axial Fourier components contribute $2r$ to Eq.~\eqref{supp:eq:alpha}, and the diagonal components contribute $r^2$, where $r=e^{-2\pi}$. Thus

\begin{equation}
 \alpha=2r+r^2<0.004004.
 \label{supp:eq:nodalgram}
\end{equation}
For the pair weight,
\begin{equation}
 w=|f|^4=
 \left(\frac32+2\cos2kx+\frac12\cos4kx\right)
 \left(\frac32+2\cos2ky+\frac12\cos4ky\right).
\end{equation}
Its constant coefficient is $w_{\bm0}=9/4$. The Gaussian factor in Eq.~\eqref{supp:eq:pairFourier} factorizes into its $x$ and $y$ parts. Writing $t=e^{-\pi}$, the total attenuated absolute Fourier sum, including the constant term, is
\begin{equation}
 \sum_{\bm q}|w_{\bm q}|e^{-q^2\ell^2/8}
 =\left(\frac32+2t+\frac12t^4\right)^2.
\end{equation}
Subtracting only the nonconstant terms therefore gives
\begin{equation}
 \kappa_{\rm Fourier}
 =\frac92-\left(\frac32+2t+\frac12t^4\right)^2.
 \label{supp:eq:nodalkappa}
\end{equation}

Taking $a=0.004004$ and $\kappa=1.98156$ in Eq.~\eqref{supp:eq:mainbound} gives
\begin{equation}
 \Delta_f\geq0.1786.
 \label{supp:eq:nodalbound}
\end{equation}
Although $f$ has nodal lines, $G$ is bounded below and $Q_f$ is coercive on $\Ran P_1$. These are the two conditions needed for the kernel and gap comparisons; a bounded ambient multiplication operator $1/f$ is not needed. The signed smooth envelope in Eq.~\eqref{supp:eq:nodal}, rather than the nonsmooth function $|f|$, is used in the contact construction.

Table~\ref{supp:tab:bounds} collects the one- and two-particle inputs used for the examples.
\begin{table}[ht]
\caption{Inputs and parent-gap bounds at $\Delta_0=1/5$. Here $\|G-\id\|\leq a$ and $g_\pm=1\pm a$. The planar rows reproduce the main-text bounds; the spherical row applies above the reference flux threshold.}
\label{supp:tab:bounds}
\begin{ruledtabular}
\begin{tabular}{llll}
Envelope & $a$ & $\kappa$ & Lower bound\\
\hline
$\epsilon=0.01,\ n=1$ & $0.002079$ & $0.9801$ & $0.1402$\\
$0\leq\epsilon\leq1/2,\ n=2$ & $0.000934$ & $0.25$ & $0.0436$\\
Nodal, Eq.~\eqref{supp:eq:nodal} & $0.004004$ & $1.98156$ & $0.1786$\\
Sphere, $\bigl||f_Q|^2-1\bigr|\leq10^{-3}$ & $0.001$ & $0.998001$ & $0.1723$
\end{tabular}
\end{ruledtabular}
\end{table}

\subsection{Spherical scalar deformations}
For a sphere of radius $R$ and flux $Q$, let $\hh_0$ be its spin-$Q/2$ LLL and let $f_Q$ be smooth. The correctly normalized projected contact is
\begin{equation}
 H_{f,Q}=c_Q\int_{S^2}\dd A\sum_{j=1}^2
   (\Psi_f\nabla_j\Psi_f)^\dagger(\Psi_f\nabla_j\Psi_f),\qquad
 c_Q=\frac{8\pi R^4(2Q-1)}{Q(Q+1)^2}.
 \label{supp:eq:spherecontact}
\end{equation}
The derivatives use orthonormal tangent directions and the monopole connection. Section~\ref{supp:app:sphere} matches this coefficient to a pair projector of energy one and proves $C_Q<8\sqrt3$. The cancellation, pair comparison, and channel argument then apply with the spherical Gram operator and coherent pair frame. In particular, $\bigl||f_Q|^2-1\bigr|\leq1/1000$ supplies the last row of Table~\ref{supp:tab:bounds}. At $Q=3(N-1)$ and sufficiently large flux, the deformed Laughlin state is the unique zero mode, and its gap is at least $0.1723$.

%% file: supplement/sections/S_geometry.tex
\section{Orbital and coordinate structure}
\label{supp:sec:geometry}
\subsection{Rank-one orbital embeddings}
Let $t(\bm r):\mathbb C\to\mathbb C^s$ be a smooth bounded column vector and let $\hh_t=t\hh_{\LLL}$ with uniformly invertible $G=P_0t^\dagger tP_0$. Consider the orbital-symmetric contact
\begin{equation}
 4\pi\ell^4\int\dd^2r\sum_{a,b=1}^s\sum_{j=x,y}
       (\Psi_a\partial_j\Psi_b)^\dagger(\Psi_a\partial_j\Psi_b).
\end{equation}
Writing $\Psi_a=t_a\chi$ gives $\Psi_a\partial_j\Psi_b=t_at_b\chi\partial_j\chi$; derivatives of $t$ multiply $\chi^2$. The scalar criterion therefore applies with Gram weight $t^\dagger t$ and pair weight $(t^\dagger t)^2$.

For $t=f n$ with $n^\dagger n=1$, this interaction is isometrically equivalent to the scalar model. One example is
$n=(\cos\theta,e^{ikx}\sin\theta)^{\mathsf T}$ with the nodal $f$ of Eq.~\eqref{supp:eq:nodal}; it has the same bound~\eqref{supp:eq:nodalbound}. The input is one field with several orbital components. For several independent input fields, derivatives of a matrix embedding generally produce terms $\chi_\mu\chi_\nu$ with $\mu\ne\nu$. Their contact channels can be studied using an appropriate pair frame, as in Sec.~\ref{supp:app:extensions}.

\subsection{A tunable chiral-graphene example}
\label{supp:app:ctbg}

Chiral twisted bilayer graphene provides a concrete setting in which an ideal Chern band carries a spatially varying layer composition. Its LLL correspondence separates this composition from the analytic factor that controls pair clustering~\cite{supp:WangZheng2021}. Here we use that separation to construct a continuum-model deformation: it changes the total spinor weight while retaining the local layer direction and an exactly flat selected band. Fix one spin, one valley and one sublattice-polarized Chern sector at a chiral magic angle. The factorization gives
\begin{equation}
 t(\bm r)=\begin{pmatrix}i\mathcal G(\bm r)\\
 \eta\mathcal G(-\bm r)\end{pmatrix},\qquad
 w(\bm r)=|\mathcal G(\bm r)|^2+|\mathcal G(-\bm r)|^2.
 \label{supp:eq:ctbgweight}
\end{equation}
The sign $\eta=\pm1$ is the intravalley inversion eigenvalue.

Here $t$ is quasiperiodic and $w$ is periodic. The effective magnetic length obeys $A_{\rm M}=2\pi\ell^2$, where $A_{\rm M}$ is the moir\'e-cell area. Multiplication by a nonzero constant allows us to choose $\langle w\rangle_{\rm cell}=1$. The zeros of either component are compatible with the orbital criterion. For the amplitude interpolation below we require that the two components have no common zero, so $w$ is strictly positive throughout the moir\'e cell.

Write the magic-angle chiral Hamiltonian, with the selected zero modes in its upper block, as
\begin{equation}
 \mathcal H_1=\begin{pmatrix}0&\mathcal D_1^\dagger\\
 \mathcal D_1&0\end{pmatrix},\qquad
 \mathcal D_1=-2iv\partial_{\bar z}\id_2+T_\alpha(\bm r),
 \qquad \partial_{\bar z}=\tfrac12(\partial_x+i\partial_y).
 \label{supp:eq:ctbgdirac}
\end{equation}
The off-diagonal layer entries of $T_\alpha$ are the chiral interlayer tunneling functions. Our complex-derivative convention is specified in Eq.~\eqref{supp:eq:ctbgdirac}; a constant energy rescaling translates to the convention of Ref.~\cite{supp:WangZheng2021}. For the positive periodic function
\begin{equation}
 a_s=w^{(s-1)/2},\qquad 0\leq s\leq1,
\end{equation}
define
\begin{align}
 \mathcal D_s&=a_s\mathcal D_1a_s^{-1}
 =\mathcal D_1+iv(s-1)(\partial_{\bar z}\ln w)\id_2,
 \label{supp:eq:ctbgmodifiedD}\\
 \mathcal H_s&=\begin{pmatrix}0&\mathcal D_s^\dagger\\
 \mathcal D_s&0\end{pmatrix}.
 \label{supp:eq:ctbgmodifiedH}
\end{align}

This is a local Hermitian single-particle Hamiltonian. To identify the physical character of the added term, write it as $v(A_{s,x}+iA_{s,y})\id_2$ in the convention of Eq.~\eqref{supp:eq:ctbgdirac}. Then
\begin{equation}
 \bm A_s=\frac{s-1}{2}\,\hat{\bm z}\mathbin{\times}\nabla\ln w,
 \qquad
 \partial_x A_{s,y}-\partial_y A_{s,x}
 =\frac{s-1}{2}\nabla^2\ln w.
 \label{supp:eq:ctbg_vector_potential}
\end{equation}
The vector potential has inverse-length units, and its curl integrates to zero over a moir\'e cell. It is diagonal in layer space but connects opposite sublattices in $\mathcal H_s$, as does a smooth modulation of intralayer bond hoppings in the leading Dirac description~\cite{supp:Vozmediano2010}. This gives the deformation a connection to graphene pseudogauge fields. Periodic pseudomagnetic fields have also been proposed as a route to nearly ideal Chern bands in strained monolayer graphene~\cite{supp:GaoStrain2023}. The specified profile defines an engineered continuum model; a microscopic strain realization would also require control of velocity and other corrections~\cite{supp:DeJuan2013}.

Only the Dirac block, rather than the full Hermitian Hamiltonian, is conjugated in Eq.~\eqref{supp:eq:ctbgmodifiedD}. The equality
$\mathcal D_s(a_s\psi)=a_s\mathcal D_1\psi$
identifies the selected kernels in both directions at every Bloch momentum. Consequently the flat band persists with wavefunctions $t_s\Phi_{\bm k}$, $t_s=a_st$. The positive periodic multiplier also preserves the Bloch boundary conditions. This construction varies a specified additional term while the interlayer tunneling remains at a magic-angle value.

For the two-layer contact in the preceding subsection,
\begin{equation}
 \Psi_{s,a}\partial_j\Psi_{s,b}
 =(t_s)_a(t_s)_b\chi_s\partial_j\chi_s,
 \qquad
 G_s=P_0w^sP_0,\qquad
 \chi_s=\sum_\lambda (G_s^{-1/2}\phi_\lambda)c_\lambda .
\end{equation}

All derivatives of $t_s$ cancel by $\chi_s^2=0$. The parent therefore depends on the total weight $t_s^\dagger t_s=w^s$. Lowering $s$ suppresses this modulation while retaining the layer texture. At $s=0$ the spinor has unit norm and the parent is isometrically equivalent to the LLL parent, so the gap criterion is positive for sufficiently small $s$. The numerical example below explores where this criterion becomes positive along the interpolation from the original chiral band at $s=1$.

\paragraph{Numerical exploration of the first magic angle.}
We also evaluated the first magic-angle solution in the standard convention $k_\theta=v=1$, with interlayer coupling $\alpha\simeq0.5857$. In a periodic gauge take
\begin{equation}
 \bm q_0=(0,-1),\quad
 \bm b_1=(\sqrt3/2,3/2),\quad
 \bm b_2=(-\sqrt3/2,3/2).
\end{equation}
The two layers use momenta $\bm k+m\bm b_1+n\bm b_2$ and
$\bm k+m\bm b_1+n\bm b_2+\bm q_0$, respectively. The three tunneling harmonics have phases $1,e^{2\pi i/3},e^{4\pi i/3}$.
We use square cutoffs $|m|,|n|\leq L$ and extract $w$ by dividing the total density by the corresponding LLL density. At the Dirac momentum the common movable zero is
$\bm r_0=(\bm a_1+2\bm a_2)/3$.
The LLL density, normalized to cell mean one, is evaluated as
\begin{equation}
 p_0(\bm r)=
 \sum_{m,n\in\mathbb Z}(-1)^{mn+m+n}
 e^{-\ell^2|\bm G_{mn}|^2/4}
 e^{i\bm G_{mn}\cdot(\bm r-\bm r_0)},\qquad
 \bm G_{mn}=m\bm b_1+n\bm b_2 .
\end{equation}
The computation retains $|m|,|n|\leq7$ in this rapidly decaying sum and samples a shifted grid to avoid dividing directly at a zero.

For the mean-normalized $v_s=w^s/\langle w^s\rangle$, write $c_{\bm G}$ and $h_{\bm G}$ for the Fourier coefficients of $v_s$ and $v_s^2$. The quantities to be bounded are
\begin{align}
 \epsilon_s&=\sum_{\bm G\ne0}|c_{\bm G}|
                          e^{-\ell^2|\bm G|^2/4},\\
 \kappa_s&\geq
 h_{\bm0}-\sum_{\bm G\ne0}|h_{\bm G}|
                          e^{-\ell^2|\bm G|^2/8}.
\end{align}
They give $g_\pm=1\pm\epsilon_s$ and
$\delta_s\leq2\epsilon_s/(1+\epsilon_s)$. A pointwise lower bound for $v_s^2$ can be used when it is stronger.

At $(L,\text{grid})=(6,128^2)$ we find sampled extrema
$w_{\min}\simeq0.3716$, $w_{\max}\simeq2.7267$. The Fourier exploration gives
\begin{center}
\begin{tabular}{cccc}
\toprule
$s$ & $\epsilon_s$ & $\kappa_s$ & gap estimate at $\Delta_0=1/5$\\
\midrule
$0$ & $0$ & $1$ & $0.2$\\
$0.02$ & $0.00433846$ & $0.9782024$ & $0.0779264$\\
$0.05$ & $0.0108955$ & $0.9455569$ & $-0.0988255$\\
$1$ & $0.2384573$ & $0.1380558$ & $-1.2711$\\
\bottomrule
\end{tabular}
\end{center}

The positive estimate at $s=0.02$ is stable across $(L,\text{grid})=(4,64^2),(5,96^2),(6,128^2)$. It illustrates how suppressing the total-weight modulation can make the criterion positive while retaining the layer texture. These are finite-cutoff numerical estimates for the continuum model.

\subsection{Changing the vortex coordinate}
The geometric description of ideal Chern bands by Estienne, Regnault, and Cr\'epel~\cite{supp:Estienne2023} involves a lowest Landau level with a nonuniform magnetic field and metric. Isothermal coordinates express its analytic structure in a complex coordinate that can differ from $x+iy$, while a scalar envelope carries the remaining amplitude variation. We first isolate the coordinate contribution to the physical contact, then allow an envelope as well. The mass-anisotropic LLL in the next subsection is a constant-metric example of this framework.

Let $F:\mathbb R^2\to\mathbb R^2$ be an orientation-preserving smooth global diffeomorphism with $J=\det DF>0$, and define
\begin{equation}
 (U_F\phi)(\bm r)=\sqrt{J(\bm r)}\,\phi(F(\bm r)).
\end{equation}
This is unitary on $L^2$. The transformed LLL is vortexable with coordinate $\zeta=F_x+iF_y$, since $U_F^\dagger\zeta U_F=z$ on the corresponding maximal domain.

Keep the physical isotropic contact in the original coordinates $\bm r$. The chain rule and $\chi^2=0$ give
\begin{equation}
 \Psi\,\partial_{r_i}\Psi
 =J\sum_j(\partial_{r_i}F_j)\chi\,\partial_{R_j}\chi,
 \qquad \bm R=F(\bm r).
\end{equation}
For an LLL field, $\chi\partial_Y\chi=i\chi\partial_X\chi$. Changing the integration measure therefore pulls the contact back to the reference pair form with weight
\begin{equation}
 w_F(\bm R)=
 \left.\frac{J(\bm r)\|DF(\bm r)\|_{\mathrm F}^2}{2}
 \right|_{\bm r=F^{-1}(\bm R)}.
 \label{supp:eq:geoweight}
\end{equation}
Here $\|\cdot\|_{\mathrm F}$ is the Frobenius matrix norm. The Jacobian survives because two projected fields supply $J^2$ and the measure supplies $J^{-1}$.

If $0<w_-\leq w_F\leq w_+<\infty$, the contact on $U_F\hh_{\LLL}$ has the pulled-back reference zero space and gap at least $w_-\Delta_0$ in every finite-particle sector.

After the unitary identification, its pair form is bounded between $w_-H_0$ and $w_+H_0$. These comparisons identify the kernel and give the stated bound on its orthogonal complement.

More generally, take the subspace $U_F(f\hh_{\LLL})$, with $f$ expressed in the reference coordinates. Unitarity of $U_F$ leaves the Gram operator equal to $G=P_0|f|^2P_0$. The same chain rule, together with the cancellation of derivatives of $f$ at particle coincidence, gives pair weight $w_F|f|^4$. The envelope criterion therefore applies with this weighted pair form in place of $Q_f$. This separates the effect of a nonuniform coordinate geometry from orbital orthogonalization without restricting $F$ to a linear squeeze.

\subsection{Mass anisotropy and the short-distance pair zero}
\label{supp:app:mass_anisotropy}
An anisotropic effective mass is a physical realization of a nonstandard vortex coordinate. For
$h_{\rm kin}=\pi_x^2/(2m_x)+\pi_y^2/(2m_y)$ in a uniform magnetic field, define
\begin{equation}
 m_c=\sqrt{m_xm_y},\qquad
 \lambda=\frac14\ln\frac{m_x}{m_y},\qquad
 (X,Y)=F_\lambda(x,y)=(e^\lambda x,e^{-\lambda}y).
 \label{supp:eq:mass_map}
\end{equation}
The map has unit Jacobian. In symmetric gauge the corresponding unitary pullback obeys
\begin{equation}
 U_\lambda^\dagger\pi_x U_\lambda=e^\lambda\Pi_X,
 \qquad
 U_\lambda^\dagger\pi_y U_\lambda=e^{-\lambda}\Pi_Y,
 \qquad
 U_\lambda^\dagger h_{\rm kin}U_\lambda
 =\frac{\Pi_X^2+\Pi_Y^2}{2m_c}.
\end{equation}
Here $\bm\Pi$ is the kinetic momentum in the reference coordinates with the same magnetic field. The magnetic length and flux density are unchanged. Consequently the anisotropic LLL is $U_\lambda\hh_{\LLL}$ and its analytic coordinate is
$\zeta_\lambda=X+iY=e^\lambda x+i e^{-\lambda}y$.
For $\lambda\ne0$, ratios of its orbitals are holomorphic in $\zeta_\lambda$ but generally not in $z=x+iy$; the change cannot be absorbed into a common scalar envelope multiplying the standard LLL.

Holding fixed the isotropic derivative contact in physical coordinates, Eq.~\eqref{supp:eq:geoweight} becomes
\begin{equation}
 w_{F_\lambda}=\cosh(2\lambda)
 =\frac{m_x+m_y}{2\sqrt{m_xm_y}}.
 \label{supp:eq:mass_contact_weight}
\end{equation}
This can also be seen directly at particle coincidence. At fixed center of mass and fixed positions of the remaining particles, analyticity and antisymmetry give the local expansion
\begin{equation}
 \psi=\mathcal E\left[
   A(\bm R,\ldots)\,\zeta_{\lambda,12}
   +B(\bm R,\ldots)\,\zeta_{\lambda,12}^{\,3}
   +\cdots\right],\qquad
 \zeta_{\lambda,12}=e^\lambda(x_1-x_2)+i e^{-\lambda}(y_1-y_2),
\end{equation}
where $\mathcal E$ is the smooth nonzero Gaussian factor. At coincidence the derivative contact probes only $A$. Derivatives of $\mathcal E$ multiply a vanishing pair wavefunction. The squared $x$ and $y$ derivatives therefore acquire factors $e^{2\lambda}$ and $e^{-2\lambda}$; dividing by their isotropic sum gives Eq.~\eqref{supp:eq:mass_contact_weight}. In field language this is the identity
$\chi\partial_Y\chi=i\chi\partial_X\chi$ used above. The cubic Laughlin zero is preserved by the squeeze.

The constant weight gives an equality of operators on each complete-plane finite-particle sector,
\begin{equation}
 U_{\lambda,N}^\dagger H_\lambda U_{\lambda,N}
 =w_{F_\lambda}H_0,\qquad
 U_{\lambda,N}=\wedge^N U_\lambda.
 \label{supp:eq:mass_spectrum}
\end{equation}
Thus every positive energy, as well as the gap, scales by $w_{F_\lambda}$; a supplied reference inequality gives
$\Delta_\lambda\geq w_{F_\lambda}\Delta_0>0$
for every finite positive mass ratio. The interaction coefficient and magnetic length are held fixed in this statement. Rescaling the interaction to keep its nonzero pair eigenvalue equal to one would instead leave the spectrum unchanged. The squeezed Laughlin polynomial uses $\prod_{i<j}(\zeta_{\lambda,i}-\zeta_{\lambda,j})^3$, with Gaussian factor $\exp[-\sum_i|\zeta_{\lambda,i}|^2/(4\ell^2)]$. It describes a correlation hole that follows the mass metric.

Metric-dependent model states and their parent Hamiltonians were constructed by Qiu \textit{et al.}~\cite{supp:Qiu2012}. Here the kinetic metric is specified as well: the squeeze changes the LLL itself while the laboratory contact remains fixed. Krishna \textit{et al.}~\cite{supp:Krishna2019} explain why the mass anisotropy is fully inherited by the many-body state in the zero-range limit. Equation~\eqref{supp:eq:mass_spectrum} supplies the corresponding spectral statement for our derivative-contact parent.

This special short-distance structure also explains the difference from the strong-anisotropy transitions found by Wang \textit{et al.}~\cite{supp:WangAnisotropy2012}. Their electrons have an isotropic LLL form factor and an anisotropic Coulomb interaction, with Fourier dependence proportional to $(q_x^2/A_c+A_cq_y^2)^{-1/2}$. It samples finite separations and can favor density-ordered states. After a squeeze it is still a finite-separation interaction, rather than a constant multiple of $V_1$. The absence of a finite-anisotropy gap closing in Eq.~\eqref{supp:eq:mass_spectrum} applies to the projected derivative contact and does not impose the same behavior on that Coulomb problem.

%% file: supplement/sections/S_domains.tex
\section{Operator definitions and domains}
\subsection{The sectorwise construction}
\label{supp:app:closure}
For fixed $N$, a bounded pair operator obeys
$\|\Ltwo(Q)|_N\|\leq\binom N2\|Q\|$. The same bound holds after the isometric one-particle dilation $J$, on the enlarged sector of total particle number at most $N$. The reference operator acts on the Fock space over $J\hh$ with spectator modes in its orthogonal complement. Its spectral inequality therefore follows from the reference sectors $0,\ldots,N$.

The analysis map $A_M$ is bounded into
$L^2(X;\Fock_{\leq N-2})$, with
$\|A_M\psi\|^2=\langle\psi,H_M\psi\rangle$.
The synthesis maps $S\otimes\id$ and $R_M$ have norms at most $\sqrt\beta$ and $\sqrt C$. Thus all terms in Eqs.~\eqref{supp:eq:noises} and \eqref{supp:eq:variance} are bounded on these fixed sectors.

For a continuous frame, exhaust its label space by finite-measure sets and first perform the calculation with the label integral restricted to such a set. The corresponding positive pair operators converge strongly to $Q$ and are bounded above by $Q$. Their pair lifts converge strongly on fixed sectors. Equivalently, the noise maps converge by applying the bounded synthesis maps to the truncated analysis map. This proves the variance identity without manipulating delta functions as Hilbert-space vectors.

The self-adjoint direct sum over finite $N$ defines the full Fock Hamiltonian. The transfer inequality is a sectorwise spectral statement with a uniform constant. As a quadratic-form inequality for $H_M^2$ it holds on $D(H_M)$, and as an operator inequality on $D(H_M^2)$.

\subsection{The maximal vortex-coordinate domain}
\label{supp:app:domain}
On ambient $L^2(\mathbb R^2)$ let $D(z)=\{\psi:z\psi\in L^2\}$ and $D_0=\hh_{\LLL}\cap D(z)$. The LLL kernel has modulus
\begin{equation}
 |P_0(z,w)|=\frac1{2\pi\ell^2}
       e^{-|z-w|^2/(4\ell^2)}.
\end{equation}
The integral kernel $(z-w)P_0(z,w)$ has both Schur integrals bounded by $2\sqrt\pi\,\ell$. Hence $[z,P_0]$ extends boundedly and $P_0$ preserves $D(z)$. This can be obtained first on a compactly supported core and then by closedness of multiplication by $z$.

Bounded scalar multiplication by $f$ and $\bar f$ preserves $D(z)$ and commutes with $z$. On $D_0$ the Gram operator satisfies
\begin{equation}
 [z,G]=[z,P_0]|f|^2\big|_{\hh_{\LLL}},
\end{equation}
so its commutator is bounded. Choose $c\geq g_+$ and write $T=\id-G/c$, $r=\|T\|<1$. For $\phi\in D_0$,
\begin{equation}
 \|zT^n\phi\|
 \leq r^n\|z\phi\|+nr^{n-1}\|[z,T]\|\,\|\phi\|.
\end{equation}
The Neumann series $G^{-1}=c^{-1}\sum_{n\geq0}T^n$ converges in the graph norm of $z$, proving that $G^{-1}$ preserves $D_0$.

If $\psi=f\phi$ lies in $\hh_f\cap D(z)$, then
$G\phi=P_0\bar f\psi\in D_0$, and therefore $\phi\in D_0$.
The reverse implication follows from boundedness of $f$. Thus
\begin{equation}
 \hh_f\cap D(z)=fD_0,\qquad z(f\phi)=f(z\phi)\in\hh_f.
\end{equation}
This argument includes nodal multipliers and does not require a pointwise inverse of $f$. For a coordinate pullback, the maximal domain is instead carried unitarily by $U_F$.

%% file: supplement/sections/S_extensions.tex
\section{More general pair frames}
\label{supp:app:extensions}
The transfer proof also handles finite decomposable expansions
\begin{equation}
 p_x=\sum_a\lambda_{x,a}u_{x,a}\wedge v_{x,a},
\end{equation}
with orthonormal vectors within each pair, by replacing the kernel in Eq.~\eqref{supp:eq:Schur} with
\begin{equation}
 k(x,y)=\sum_{a,b}|\lambda_{x,a}||\lambda_{y,b}|
         \|U_{x,a}^\dagger U_{y,b}\|_1.
\end{equation}
The creation amplitudes in $R_M$ carry $\lambda_{x,a}$, while $B_x^M$ carries its conjugate. The two-frame bound~\eqref{supp:eq:overlaplemma} and the triangle inequality give the same Schur bound. The second noise term depends on $\beta=\|Q\|$, so a nonprojection frame is included without further change.

Several contact channels can be represented by adjoining a channel label to $x$. The overlap condition then includes the cross-channel blocks. Nonorthogonal spanning pairs can be orthonormalized with their wedge norm absorbed into $\lambda$. These statements provide operational one-particle conditions for the generalization.

As a simple explicit family, take disjoint four-mode cells and
\begin{equation}
 p_j=\frac{e_{j1}\wedge e_{j2}+e_{j3}\wedge e_{j4}}{\sqrt2}.
\end{equation}
Here $\beta=1$ and $C=2$, because overlaps between cells vanish. The local operator $B(p_j)^\dagger B(p_j)$ has energies $0,1/2,1$: its three-particle sector is $(1/2)\id$ and its occupied four-particle state has energy one. Even operators on different cells commute, giving a reference gap $1/2$ for any number of cells. Equation~\eqref{supp:eq:transfer} therefore gives
\begin{equation}
 \gap H_M\geq\frac12-2\delta_M-\delta_M^2
\end{equation}
when the right-hand side is positive. This illustrates the finite-Slater-rank extension independently of Landau-level geometry.

%% file: main.bbl
\begin{thebibliography}{36}\makeatletter
\providecommand \@ifxundefined [1]{\@ifx{#1\undefined}
}\providecommand \@ifnum [1]{\ifnum #1\expandafter \@firstoftwo
 \else \expandafter \@secondoftwo
 \fi
}\providecommand \@ifx [1]{\ifx #1\expandafter \@firstoftwo
 \else \expandafter \@secondoftwo
 \fi
}\providecommand \natexlab [1]{#1}\providecommand \enquote  [1]{``#1''}\providecommand \bibnamefont  [1]{#1}\providecommand \bibfnamefont [1]{#1}\providecommand \citenamefont [1]{#1}\providecommand \href@noop [0]{\@secondoftwo}\providecommand \href [0]{\begingroup \@sanitize@url \@href}\providecommand \@href[1]{\@@startlink{#1}\@@href}\providecommand \@@href[1]{\endgroup#1\@@endlink}\providecommand \@sanitize@url [0]{\catcode `\\12\catcode `\$12\catcode
  `\&12\catcode `\#12\catcode `\^12\catcode `\_12\catcode `\%12\relax}\providecommand \@@startlink[1]{}\providecommand \@@endlink[0]{}\providecommand \url  [0]{\begingroup\@sanitize@url \@url }\providecommand \@url [1]{\endgroup\@href {#1}{\urlprefix }}\providecommand \urlprefix  [0]{URL }\providecommand \Eprint [0]{\href }\providecommand \doibase [0]{https://doi.org/}\providecommand \selectlanguage [0]{\@gobble}\providecommand \bibinfo  [0]{\@secondoftwo}\providecommand \bibfield  [0]{\@secondoftwo}\providecommand \translation [1]{[#1]}\providecommand \BibitemOpen [0]{}\providecommand \bibitemStop [0]{}\providecommand \bibitemNoStop [0]{.\EOS\space}\providecommand \EOS [0]{\spacefactor3000\relax}\providecommand \BibitemShut  [1]{\csname bibitem#1\endcsname}\let\auto@bib@innerbib\@empty
\bibitem [{\citenamefont {Tsui}\ \emph {et~al.}(1982)\citenamefont {Tsui},
  \citenamefont {Stormer},\ and\ \citenamefont {Gossard}}]{Tsui1982}\BibitemOpen
  \bibfield  {author} {\bibinfo {author} {\bibfnamefont {D.~C.}\ \bibnamefont
  {Tsui}}, \bibinfo {author} {\bibfnamefont {H.~L.}\ \bibnamefont {Stormer}},\
  and\ \bibinfo {author} {\bibfnamefont {A.~C.}\ \bibnamefont {Gossard}},\
  }\bibfield  {title} {\bibinfo {title} {Two-dimensional magnetotransport in
  the extreme quantum limit},\ }\href
  {https://doi.org/10.1103/PhysRevLett.48.1559} {\bibfield  {journal} {\bibinfo
   {journal} {Phys. Rev. Lett.}\ }\textbf {\bibinfo {volume} {48}},\ \bibinfo
  {pages} {1559} (\bibinfo {year} {1982})}\BibitemShut {NoStop}\bibitem [{\citenamefont {Laughlin}(1983)}]{Laughlin1983}\BibitemOpen
  \bibfield  {author} {\bibinfo {author} {\bibfnamefont {R.~B.}\ \bibnamefont
  {Laughlin}},\ }\bibfield  {title} {\bibinfo {title} {Anomalous quantum {Hall}
  effect: An incompressible quantum fluid with fractionally charged
  excitations},\ }\href {https://doi.org/10.1103/PhysRevLett.50.1395}
  {\bibfield  {journal} {\bibinfo  {journal} {Phys. Rev. Lett.}\ }\textbf
  {\bibinfo {volume} {50}},\ \bibinfo {pages} {1395} (\bibinfo {year}
  {1983})}\BibitemShut {NoStop}\bibitem [{\citenamefont {Arovas}\ \emph {et~al.}(1984)\citenamefont {Arovas},
  \citenamefont {Schrieffer},\ and\ \citenamefont {Wilczek}}]{Arovas1984}\BibitemOpen
  \bibfield  {author} {\bibinfo {author} {\bibfnamefont {D.}~\bibnamefont
  {Arovas}}, \bibinfo {author} {\bibfnamefont {J.~R.}\ \bibnamefont
  {Schrieffer}},\ and\ \bibinfo {author} {\bibfnamefont {F.}~\bibnamefont
  {Wilczek}},\ }\bibfield  {title} {\bibinfo {title} {Fractional statistics and
  the quantum {Hall} effect},\ }\href
  {https://doi.org/10.1103/PhysRevLett.53.722} {\bibfield  {journal} {\bibinfo
  {journal} {Phys. Rev. Lett.}\ }\textbf {\bibinfo {volume} {53}},\ \bibinfo
  {pages} {722} (\bibinfo {year} {1984})}\BibitemShut {NoStop}\bibitem [{\citenamefont {Haldane}(1983)}]{Haldane1983}\BibitemOpen
  \bibfield  {author} {\bibinfo {author} {\bibfnamefont {F.~D.~M.}\
  \bibnamefont {Haldane}},\ }\bibfield  {title} {\bibinfo {title} {Fractional
  quantization of the {Hall} effect: A hierarchy of incompressible quantum
  fluid states},\ }\href {https://doi.org/10.1103/PhysRevLett.51.605}
  {\bibfield  {journal} {\bibinfo  {journal} {Phys. Rev. Lett.}\ }\textbf
  {\bibinfo {volume} {51}},\ \bibinfo {pages} {605} (\bibinfo {year}
  {1983})}\BibitemShut {NoStop}\bibitem [{\citenamefont {Trugman}\ and\ \citenamefont
  {Kivelson}(1985)}]{Trugman1985}\BibitemOpen
  \bibfield  {author} {\bibinfo {author} {\bibfnamefont {S.~A.}\ \bibnamefont
  {Trugman}}\ and\ \bibinfo {author} {\bibfnamefont {S.}~\bibnamefont
  {Kivelson}},\ }\bibfield  {title} {\bibinfo {title} {Exact results for the
  fractional quantum {Hall} effect with general interactions},\ }\href
  {https://doi.org/10.1103/PhysRevB.31.5280} {\bibfield  {journal} {\bibinfo
  {journal} {Phys. Rev. B}\ }\textbf {\bibinfo {volume} {31}},\ \bibinfo
  {pages} {5280} (\bibinfo {year} {1985})}\BibitemShut {NoStop}\bibitem [{\citenamefont {Tang}\ \emph {et~al.}(2011)\citenamefont {Tang},
  \citenamefont {Mei},\ and\ \citenamefont {Wen}}]{Tang2011}\BibitemOpen
  \bibfield  {author} {\bibinfo {author} {\bibfnamefont {E.}~\bibnamefont
  {Tang}}, \bibinfo {author} {\bibfnamefont {J.-W.}\ \bibnamefont {Mei}},\ and\
  \bibinfo {author} {\bibfnamefont {X.-G.}\ \bibnamefont {Wen}},\ }\bibfield
  {title} {\bibinfo {title} {High-temperature fractional quantum {Hall}
  states},\ }\href {https://doi.org/10.1103/PhysRevLett.106.236802} {\bibfield
  {journal} {\bibinfo  {journal} {Phys. Rev. Lett.}\ }\textbf {\bibinfo
  {volume} {106}},\ \bibinfo {pages} {236802} (\bibinfo {year}
  {2011})}\BibitemShut {NoStop}\bibitem [{\citenamefont {Sun}\ \emph {et~al.}(2011)\citenamefont {Sun},
  \citenamefont {Gu}, \citenamefont {Katsura},\ and\ \citenamefont
  {Das~Sarma}}]{Sun2011}\BibitemOpen
  \bibfield  {author} {\bibinfo {author} {\bibfnamefont {K.}~\bibnamefont
  {Sun}}, \bibinfo {author} {\bibfnamefont {Z.}~\bibnamefont {Gu}}, \bibinfo
  {author} {\bibfnamefont {H.}~\bibnamefont {Katsura}},\ and\ \bibinfo {author}
  {\bibfnamefont {S.}~\bibnamefont {Das~Sarma}},\ }\bibfield  {title} {\bibinfo
  {title} {Nearly flatbands with nontrivial topology},\ }\href
  {https://doi.org/10.1103/PhysRevLett.106.236803} {\bibfield  {journal}
  {\bibinfo  {journal} {Phys. Rev. Lett.}\ }\textbf {\bibinfo {volume} {106}},\
  \bibinfo {pages} {236803} (\bibinfo {year} {2011})}\BibitemShut {NoStop}\bibitem [{\citenamefont {Neupert}\ \emph {et~al.}(2011)\citenamefont
  {Neupert}, \citenamefont {Santos}, \citenamefont {Chamon},\ and\
  \citenamefont {Mudry}}]{Neupert2011}\BibitemOpen
  \bibfield  {author} {\bibinfo {author} {\bibfnamefont {T.}~\bibnamefont
  {Neupert}}, \bibinfo {author} {\bibfnamefont {L.}~\bibnamefont {Santos}},
  \bibinfo {author} {\bibfnamefont {C.}~\bibnamefont {Chamon}},\ and\ \bibinfo
  {author} {\bibfnamefont {C.}~\bibnamefont {Mudry}},\ }\bibfield  {title}
  {\bibinfo {title} {Fractional quantum {Hall} states at zero magnetic field},\
  }\href {https://doi.org/10.1103/PhysRevLett.106.236804} {\bibfield  {journal}
  {\bibinfo  {journal} {Phys. Rev. Lett.}\ }\textbf {\bibinfo {volume} {106}},\
  \bibinfo {pages} {236804} (\bibinfo {year} {2011})}\BibitemShut {NoStop}\bibitem [{\citenamefont {Sheng}\ \emph {et~al.}(2011)\citenamefont {Sheng},
  \citenamefont {Gu}, \citenamefont {Sun},\ and\ \citenamefont
  {Sheng}}]{Sheng2011}\BibitemOpen
  \bibfield  {author} {\bibinfo {author} {\bibfnamefont {D.~N.}\ \bibnamefont
  {Sheng}}, \bibinfo {author} {\bibfnamefont {Z.-C.}\ \bibnamefont {Gu}},
  \bibinfo {author} {\bibfnamefont {K.}~\bibnamefont {Sun}},\ and\ \bibinfo
  {author} {\bibfnamefont {L.}~\bibnamefont {Sheng}},\ }\bibfield  {title}
  {\bibinfo {title} {Fractional quantum {Hall} effect in the absence of
  {Landau} levels},\ }\href {https://doi.org/10.1038/ncomms1380} {\bibfield
  {journal} {\bibinfo  {journal} {Nat. Commun.}\ }\textbf {\bibinfo {volume}
  {2}},\ \bibinfo {pages} {389} (\bibinfo {year} {2011})}\BibitemShut {NoStop}\bibitem [{\citenamefont {Regnault}\ and\ \citenamefont
  {Bernevig}(2011)}]{Regnault2011}\BibitemOpen
  \bibfield  {author} {\bibinfo {author} {\bibfnamefont {N.}~\bibnamefont
  {Regnault}}\ and\ \bibinfo {author} {\bibfnamefont {B.~A.}\ \bibnamefont
  {Bernevig}},\ }\bibfield  {title} {\bibinfo {title} {Fractional {Chern}
  insulator},\ }\href {https://doi.org/10.1103/PhysRevX.1.021014} {\bibfield
  {journal} {\bibinfo  {journal} {Phys. Rev. X}\ }\textbf {\bibinfo {volume}
  {1}},\ \bibinfo {pages} {021014} (\bibinfo {year} {2011})}\BibitemShut
  {NoStop}\bibitem [{\citenamefont {Roy}(2014)}]{Roy2014}\BibitemOpen
  \bibfield  {author} {\bibinfo {author} {\bibfnamefont {R.}~\bibnamefont
  {Roy}},\ }\bibfield  {title} {\bibinfo {title} {Band geometry of fractional
  topological insulators},\ }\href {https://doi.org/10.1103/PhysRevB.90.165139}
  {\bibfield  {journal} {\bibinfo  {journal} {Phys. Rev. B}\ }\textbf {\bibinfo
  {volume} {90}},\ \bibinfo {pages} {165139} (\bibinfo {year}
  {2014})}\BibitemShut {NoStop}\bibitem [{\citenamefont {Wang}\ \emph
  {et~al.}(2021{\natexlab{a}})\citenamefont {Wang}, \citenamefont {Cano},
  \citenamefont {Millis}, \citenamefont {Liu},\ and\ \citenamefont
  {Yang}}]{Wang2021}\BibitemOpen
  \bibfield  {author} {\bibinfo {author} {\bibfnamefont {J.}~\bibnamefont
  {Wang}}, \bibinfo {author} {\bibfnamefont {J.}~\bibnamefont {Cano}}, \bibinfo
  {author} {\bibfnamefont {A.~J.}\ \bibnamefont {Millis}}, \bibinfo {author}
  {\bibfnamefont {Z.}~\bibnamefont {Liu}},\ and\ \bibinfo {author}
  {\bibfnamefont {B.}~\bibnamefont {Yang}},\ }\bibfield  {title} {\bibinfo
  {title} {Exact {Landau} level description of geometry and interaction in a
  flatband},\ }\href {https://doi.org/10.1103/PhysRevLett.127.246403}
  {\bibfield  {journal} {\bibinfo  {journal} {Phys. Rev. Lett.}\ }\textbf
  {\bibinfo {volume} {127}},\ \bibinfo {pages} {246403} (\bibinfo {year}
  {2021}{\natexlab{a}})}\BibitemShut {NoStop}\bibitem [{\citenamefont {Wang}\ \emph {et~al.}(2023)\citenamefont {Wang},
  \citenamefont {Klevtsov},\ and\ \citenamefont {Liu}}]{Wang2023}\BibitemOpen
  \bibfield  {author} {\bibinfo {author} {\bibfnamefont {J.}~\bibnamefont
  {Wang}}, \bibinfo {author} {\bibfnamefont {S.}~\bibnamefont {Klevtsov}},\
  and\ \bibinfo {author} {\bibfnamefont {Z.}~\bibnamefont {Liu}},\ }\bibfield
  {title} {\bibinfo {title} {Origin of model fractional {Chern} insulators in
  all topological ideal flatbands: Explicit color-entangled wave function and
  exact density algebra},\ }\href
  {https://doi.org/10.1103/PhysRevResearch.5.023167} {\bibfield  {journal}
  {\bibinfo  {journal} {Phys. Rev. Research}\ }\textbf {\bibinfo {volume}
  {5}},\ \bibinfo {pages} {023167} (\bibinfo {year} {2023})}\BibitemShut
  {NoStop}\bibitem [{\citenamefont {Cano}\ and\ \citenamefont
  {Wang}(2026)}]{CanoWang2026}\BibitemOpen
  \bibfield  {author} {\bibinfo {author} {\bibfnamefont {J.}~\bibnamefont
  {Cano}}\ and\ \bibinfo {author} {\bibfnamefont {J.}~\bibnamefont {Wang}},\
  }\bibfield  {title} {\bibinfo {title} {Ideal quantum geometry for fractional
  {Chern} insulators},\ }\href
  {https://doi.org/10.1146/annurev-conmatphys-032725-123642} {\bibfield
  {journal} {\bibinfo  {journal} {Annu. Rev. Condens. Matter Phys.}\ }\textbf
  {\bibinfo {volume} {18}},\ \bibinfo {pages} {in press} (\bibinfo {year}
  {2026})}\BibitemShut {NoStop}\bibitem [{\citenamefont {Ledwith}\ \emph {et~al.}(2023)\citenamefont
  {Ledwith}, \citenamefont {Vishwanath},\ and\ \citenamefont
  {Parker}}]{Ledwith2023}\BibitemOpen
  \bibfield  {author} {\bibinfo {author} {\bibfnamefont {P.~J.}\ \bibnamefont
  {Ledwith}}, \bibinfo {author} {\bibfnamefont {A.}~\bibnamefont
  {Vishwanath}},\ and\ \bibinfo {author} {\bibfnamefont {D.~E.}\ \bibnamefont
  {Parker}},\ }\bibfield  {title} {\bibinfo {title} {Vortexability: A unifying
  criterion for ideal fractional {Chern} insulators},\ }\href
  {https://doi.org/10.1103/PhysRevB.108.205144} {\bibfield  {journal} {\bibinfo
   {journal} {Phys. Rev. B}\ }\textbf {\bibinfo {volume} {108}},\ \bibinfo
  {pages} {205144} (\bibinfo {year} {2023})}\BibitemShut {NoStop}\bibitem [{\citenamefont {Ledwith}\ \emph {et~al.}(2020)\citenamefont
  {Ledwith}, \citenamefont {Tarnopolsky}, \citenamefont {Khalaf},\ and\
  \citenamefont {Vishwanath}}]{Ledwith2020}\BibitemOpen
  \bibfield  {author} {\bibinfo {author} {\bibfnamefont {P.~J.}\ \bibnamefont
  {Ledwith}}, \bibinfo {author} {\bibfnamefont {G.}~\bibnamefont
  {Tarnopolsky}}, \bibinfo {author} {\bibfnamefont {E.}~\bibnamefont
  {Khalaf}},\ and\ \bibinfo {author} {\bibfnamefont {A.}~\bibnamefont
  {Vishwanath}},\ }\bibfield  {title} {\bibinfo {title} {Fractional {Chern}
  insulator states in twisted bilayer graphene: An analytical approach},\
  }\href {https://doi.org/10.1103/PhysRevResearch.2.023237} {\bibfield
  {journal} {\bibinfo  {journal} {Phys. Rev. Research}\ }\textbf {\bibinfo
  {volume} {2}},\ \bibinfo {pages} {023237} (\bibinfo {year}
  {2020})}\BibitemShut {NoStop}\bibitem [{\citenamefont {Estienne}\ \emph {et~al.}(2023)\citenamefont
  {Estienne}, \citenamefont {Regnault},\ and\ \citenamefont
  {Cr{\'e}pel}}]{Estienne2023}\BibitemOpen
  \bibfield  {author} {\bibinfo {author} {\bibfnamefont {B.}~\bibnamefont
  {Estienne}}, \bibinfo {author} {\bibfnamefont {N.}~\bibnamefont {Regnault}},\
  and\ \bibinfo {author} {\bibfnamefont {V.}~\bibnamefont {Cr{\'e}pel}},\
  }\bibfield  {title} {\bibinfo {title} {Ideal {Chern} bands are {Landau}
  levels in curved space},\ }\href
  {https://doi.org/10.1103/PhysRevResearch.5.L032048} {\bibfield  {journal}
  {\bibinfo  {journal} {Phys. Rev. Research}\ }\textbf {\bibinfo {volume}
  {5}},\ \bibinfo {pages} {L032048} (\bibinfo {year} {2023})}\BibitemShut
  {NoStop}\bibitem [{\citenamefont {Tarnopolsky}\ \emph {et~al.}(2019)\citenamefont
  {Tarnopolsky}, \citenamefont {Kruchkov},\ and\ \citenamefont
  {Vishwanath}}]{Tarnopolsky2019}\BibitemOpen
  \bibfield  {author} {\bibinfo {author} {\bibfnamefont {G.}~\bibnamefont
  {Tarnopolsky}}, \bibinfo {author} {\bibfnamefont {A.~J.}\ \bibnamefont
  {Kruchkov}},\ and\ \bibinfo {author} {\bibfnamefont {A.}~\bibnamefont
  {Vishwanath}},\ }\bibfield  {title} {\bibinfo {title} {Origin of magic angles
  in twisted bilayer graphene},\ }\href
  {https://doi.org/10.1103/PhysRevLett.122.106405} {\bibfield  {journal}
  {\bibinfo  {journal} {Phys. Rev. Lett.}\ }\textbf {\bibinfo {volume} {122}},\
  \bibinfo {pages} {106405} (\bibinfo {year} {2019})}\BibitemShut {NoStop}\bibitem [{\citenamefont {Aharonov}\ and\ \citenamefont
  {Casher}(1979)}]{Aharonov1979}\BibitemOpen
  \bibfield  {author} {\bibinfo {author} {\bibfnamefont {Y.}~\bibnamefont
  {Aharonov}}\ and\ \bibinfo {author} {\bibfnamefont {A.}~\bibnamefont
  {Casher}},\ }\bibfield  {title} {\bibinfo {title} {Ground state of a
  spin-$\frac12$ charged particle in a two-dimensional magnetic field},\ }\href
  {https://doi.org/10.1103/PhysRevA.19.2461} {\bibfield  {journal} {\bibinfo
  {journal} {Phys. Rev. A}\ }\textbf {\bibinfo {volume} {19}},\ \bibinfo
  {pages} {2461} (\bibinfo {year} {1979})}\BibitemShut {NoStop}\bibitem [{\citenamefont {Morales-Dur{\'a}n}\ \emph {et~al.}(2024)\citenamefont
  {Morales-Dur{\'a}n}, \citenamefont {Wei}, \citenamefont {Shi},\ and\
  \citenamefont {MacDonald}}]{MoralesDuran2024}\BibitemOpen
  \bibfield  {author} {\bibinfo {author} {\bibfnamefont {N.}~\bibnamefont
  {Morales-Dur{\'a}n}}, \bibinfo {author} {\bibfnamefont {N.}~\bibnamefont
  {Wei}}, \bibinfo {author} {\bibfnamefont {J.}~\bibnamefont {Shi}},\ and\
  \bibinfo {author} {\bibfnamefont {A.~H.}\ \bibnamefont {MacDonald}},\
  }\bibfield  {title} {\bibinfo {title} {Magic angles and fractional {Chern}
  insulators in twisted homobilayer transition metal dichalcogenides},\ }\href
  {https://doi.org/10.1103/PhysRevLett.132.096602} {\bibfield  {journal}
  {\bibinfo  {journal} {Phys. Rev. Lett.}\ }\textbf {\bibinfo {volume} {132}},\
  \bibinfo {pages} {096602} (\bibinfo {year} {2024})}\BibitemShut {NoStop}\bibitem [{\citenamefont {Shi}\ \emph {et~al.}(2024)\citenamefont {Shi},
  \citenamefont {Morales-Dur{\'a}n}, \citenamefont {Khalaf},\ and\
  \citenamefont {MacDonald}}]{Shi2024}\BibitemOpen
  \bibfield  {author} {\bibinfo {author} {\bibfnamefont {J.}~\bibnamefont
  {Shi}}, \bibinfo {author} {\bibfnamefont {N.}~\bibnamefont
  {Morales-Dur{\'a}n}}, \bibinfo {author} {\bibfnamefont {E.}~\bibnamefont
  {Khalaf}},\ and\ \bibinfo {author} {\bibfnamefont {A.~H.}\ \bibnamefont
  {MacDonald}},\ }\bibfield  {title} {\bibinfo {title} {Adiabatic approximation
  and {Aharonov--Casher} bands in twisted homobilayer transition metal
  dichalcogenides},\ }\href {https://doi.org/10.1103/PhysRevB.110.035130}
  {\bibfield  {journal} {\bibinfo  {journal} {Phys. Rev. B}\ }\textbf {\bibinfo
  {volume} {110}},\ \bibinfo {pages} {035130} (\bibinfo {year}
  {2024})}\BibitemShut {NoStop}\bibitem [{\citenamefont {Cai}\ \emph {et~al.}(2023)\citenamefont {Cai},
  \citenamefont {Anderson}, \citenamefont {Wang}, \citenamefont {Zhang},
  \citenamefont {Liu}, \citenamefont {Holtzmann}, \citenamefont {Zhang},
  \citenamefont {Fan}, \citenamefont {Taniguchi}, \citenamefont {Watanabe},
  \citenamefont {Ran}, \citenamefont {Cao}, \citenamefont {Fu}, \citenamefont
  {Xiao}, \citenamefont {Yao},\ and\ \citenamefont {Xu}}]{Cai2023}\BibitemOpen
  \bibfield  {author} {\bibinfo {author} {\bibfnamefont {J.}~\bibnamefont
  {Cai}}, \bibinfo {author} {\bibfnamefont {E.}~\bibnamefont {Anderson}},
  \bibinfo {author} {\bibfnamefont {C.}~\bibnamefont {Wang}}, \bibinfo {author}
  {\bibfnamefont {X.}~\bibnamefont {Zhang}}, \bibinfo {author} {\bibfnamefont
  {X.}~\bibnamefont {Liu}}, \bibinfo {author} {\bibfnamefont {W.}~\bibnamefont
  {Holtzmann}}, \bibinfo {author} {\bibfnamefont {Y.}~\bibnamefont {Zhang}},
  \bibinfo {author} {\bibfnamefont {F.}~\bibnamefont {Fan}}, \bibinfo {author}
  {\bibfnamefont {T.}~\bibnamefont {Taniguchi}}, \bibinfo {author}
  {\bibfnamefont {K.}~\bibnamefont {Watanabe}}, \bibinfo {author}
  {\bibfnamefont {Y.}~\bibnamefont {Ran}}, \bibinfo {author} {\bibfnamefont
  {T.}~\bibnamefont {Cao}}, \bibinfo {author} {\bibfnamefont {L.}~\bibnamefont
  {Fu}}, \bibinfo {author} {\bibfnamefont {D.}~\bibnamefont {Xiao}}, \bibinfo
  {author} {\bibfnamefont {W.}~\bibnamefont {Yao}},\ and\ \bibinfo {author}
  {\bibfnamefont {X.}~\bibnamefont {Xu}},\ }\bibfield  {title} {\bibinfo
  {title} {Signatures of fractional quantum anomalous {Hall} states in twisted
  {MoTe}$_2$},\ }\href {https://doi.org/10.1038/s41586-023-06289-w} {\bibfield
  {journal} {\bibinfo  {journal} {Nature}\ }\textbf {\bibinfo {volume} {622}},\
  \bibinfo {pages} {63} (\bibinfo {year} {2023})}\BibitemShut {NoStop}\bibitem [{\citenamefont {Zeng}\ \emph {et~al.}(2023)\citenamefont {Zeng},
  \citenamefont {Xia}, \citenamefont {Kang}, \citenamefont {Zhu}, \citenamefont
  {Kn{\"u}ppel}, \citenamefont {Vaswani}, \citenamefont {Watanabe},
  \citenamefont {Taniguchi}, \citenamefont {Mak},\ and\ \citenamefont
  {Shan}}]{Zeng2023}\BibitemOpen
  \bibfield  {author} {\bibinfo {author} {\bibfnamefont {Y.}~\bibnamefont
  {Zeng}}, \bibinfo {author} {\bibfnamefont {Z.}~\bibnamefont {Xia}}, \bibinfo
  {author} {\bibfnamefont {K.}~\bibnamefont {Kang}}, \bibinfo {author}
  {\bibfnamefont {J.}~\bibnamefont {Zhu}}, \bibinfo {author} {\bibfnamefont
  {P.}~\bibnamefont {Kn{\"u}ppel}}, \bibinfo {author} {\bibfnamefont
  {C.}~\bibnamefont {Vaswani}}, \bibinfo {author} {\bibfnamefont
  {K.}~\bibnamefont {Watanabe}}, \bibinfo {author} {\bibfnamefont
  {T.}~\bibnamefont {Taniguchi}}, \bibinfo {author} {\bibfnamefont {K.~F.}\
  \bibnamefont {Mak}},\ and\ \bibinfo {author} {\bibfnamefont {J.}~\bibnamefont
  {Shan}},\ }\bibfield  {title} {\bibinfo {title} {Thermodynamic evidence of
  fractional {Chern} insulator in moir{\'e} {MoTe}$_2$},\ }\href
  {https://doi.org/10.1038/s41586-023-06452-3} {\bibfield  {journal} {\bibinfo
  {journal} {Nature}\ }\textbf {\bibinfo {volume} {622}},\ \bibinfo {pages}
  {69} (\bibinfo {year} {2023})}\BibitemShut {NoStop}\bibitem [{\citenamefont {Park}\ \emph {et~al.}(2023)\citenamefont {Park},
  \citenamefont {Cai}, \citenamefont {Anderson}, \citenamefont {Zhang},
  \citenamefont {Zhu}, \citenamefont {Liu}, \citenamefont {Wang}, \citenamefont
  {Holtzmann}, \citenamefont {Hu}, \citenamefont {Liu}, \citenamefont
  {Taniguchi}, \citenamefont {Watanabe}, \citenamefont {Chu}, \citenamefont
  {Cao}, \citenamefont {Fu}, \citenamefont {Yao}, \citenamefont {Chang},
  \citenamefont {Cobden}, \citenamefont {Xiao},\ and\ \citenamefont
  {Xu}}]{Park2023}\BibitemOpen
  \bibfield  {author} {\bibinfo {author} {\bibfnamefont {H.}~\bibnamefont
  {Park}}, \bibinfo {author} {\bibfnamefont {J.}~\bibnamefont {Cai}}, \bibinfo
  {author} {\bibfnamefont {E.}~\bibnamefont {Anderson}}, \bibinfo {author}
  {\bibfnamefont {Y.}~\bibnamefont {Zhang}}, \bibinfo {author} {\bibfnamefont
  {J.}~\bibnamefont {Zhu}}, \bibinfo {author} {\bibfnamefont {X.}~\bibnamefont
  {Liu}}, \bibinfo {author} {\bibfnamefont {C.}~\bibnamefont {Wang}}, \bibinfo
  {author} {\bibfnamefont {W.}~\bibnamefont {Holtzmann}}, \bibinfo {author}
  {\bibfnamefont {C.}~\bibnamefont {Hu}}, \bibinfo {author} {\bibfnamefont
  {Z.}~\bibnamefont {Liu}}, \bibinfo {author} {\bibfnamefont {T.}~\bibnamefont
  {Taniguchi}}, \bibinfo {author} {\bibfnamefont {K.}~\bibnamefont {Watanabe}},
  \bibinfo {author} {\bibfnamefont {J.-H.}\ \bibnamefont {Chu}}, \bibinfo
  {author} {\bibfnamefont {T.}~\bibnamefont {Cao}}, \bibinfo {author}
  {\bibfnamefont {L.}~\bibnamefont {Fu}}, \bibinfo {author} {\bibfnamefont
  {W.}~\bibnamefont {Yao}}, \bibinfo {author} {\bibfnamefont {C.-Z.}\
  \bibnamefont {Chang}}, \bibinfo {author} {\bibfnamefont {D.}~\bibnamefont
  {Cobden}}, \bibinfo {author} {\bibfnamefont {D.}~\bibnamefont {Xiao}},\ and\
  \bibinfo {author} {\bibfnamefont {X.}~\bibnamefont {Xu}},\ }\bibfield
  {title} {\bibinfo {title} {Observation of fractionally quantized anomalous
  {Hall} effect},\ }\href {https://doi.org/10.1038/s41586-023-06536-0}
  {\bibfield  {journal} {\bibinfo  {journal} {Nature}\ }\textbf {\bibinfo
  {volume} {622}},\ \bibinfo {pages} {74} (\bibinfo {year} {2023})}\BibitemShut
  {NoStop}\bibitem [{\citenamefont {Xu}\ \emph {et~al.}(2023)\citenamefont {Xu},
  \citenamefont {Sun}, \citenamefont {Jia}, \citenamefont {Liu}, \citenamefont
  {Xu}, \citenamefont {Li}, \citenamefont {Gu}, \citenamefont {Watanabe},
  \citenamefont {Taniguchi}, \citenamefont {Tong}, \citenamefont {Jia},
  \citenamefont {Shi}, \citenamefont {Jiang}, \citenamefont {Zhang},
  \citenamefont {Liu},\ and\ \citenamefont {Li}}]{Xu2023}\BibitemOpen
  \bibfield  {author} {\bibinfo {author} {\bibfnamefont {F.}~\bibnamefont
  {Xu}}, \bibinfo {author} {\bibfnamefont {Z.}~\bibnamefont {Sun}}, \bibinfo
  {author} {\bibfnamefont {T.}~\bibnamefont {Jia}}, \bibinfo {author}
  {\bibfnamefont {C.}~\bibnamefont {Liu}}, \bibinfo {author} {\bibfnamefont
  {C.}~\bibnamefont {Xu}}, \bibinfo {author} {\bibfnamefont {C.}~\bibnamefont
  {Li}}, \bibinfo {author} {\bibfnamefont {Y.}~\bibnamefont {Gu}}, \bibinfo
  {author} {\bibfnamefont {K.}~\bibnamefont {Watanabe}}, \bibinfo {author}
  {\bibfnamefont {T.}~\bibnamefont {Taniguchi}}, \bibinfo {author}
  {\bibfnamefont {B.}~\bibnamefont {Tong}}, \bibinfo {author} {\bibfnamefont
  {J.}~\bibnamefont {Jia}}, \bibinfo {author} {\bibfnamefont {Z.}~\bibnamefont
  {Shi}}, \bibinfo {author} {\bibfnamefont {S.}~\bibnamefont {Jiang}}, \bibinfo
  {author} {\bibfnamefont {Y.}~\bibnamefont {Zhang}}, \bibinfo {author}
  {\bibfnamefont {X.}~\bibnamefont {Liu}},\ and\ \bibinfo {author}
  {\bibfnamefont {T.}~\bibnamefont {Li}},\ }\bibfield  {title} {\bibinfo
  {title} {Observation of integer and fractional quantum anomalous {Hall}
  effects in twisted bilayer {MoTe}$_2$},\ }\href
  {https://doi.org/10.1103/PhysRevX.13.031037} {\bibfield  {journal} {\bibinfo
  {journal} {Phys. Rev. X}\ }\textbf {\bibinfo {volume} {13}},\ \bibinfo
  {pages} {031037} (\bibinfo {year} {2023})}\BibitemShut {NoStop}\bibitem [{\citenamefont {Liu}\ \emph {et~al.}(2025)\citenamefont {Liu},
  \citenamefont {Yang}, \citenamefont {Abouelkomsan}, \citenamefont {Liu},\
  and\ \citenamefont {Bergholtz}}]{Liu2025}\BibitemOpen
  \bibfield  {author} {\bibinfo {author} {\bibfnamefont {H.}~\bibnamefont
  {Liu}}, \bibinfo {author} {\bibfnamefont {K.}~\bibnamefont {Yang}}, \bibinfo
  {author} {\bibfnamefont {A.}~\bibnamefont {Abouelkomsan}}, \bibinfo {author}
  {\bibfnamefont {Z.}~\bibnamefont {Liu}},\ and\ \bibinfo {author}
  {\bibfnamefont {E.~J.}\ \bibnamefont {Bergholtz}},\ }\bibfield  {title}
  {\bibinfo {title} {Broken symmetry in ideal {Chern} bands},\ }\href
  {https://doi.org/10.1103/PhysRevB.111.L201105} {\bibfield  {journal}
  {\bibinfo  {journal} {Phys. Rev. B}\ }\textbf {\bibinfo {volume} {111}},\
  \bibinfo {pages} {L201105} (\bibinfo {year} {2025})}\BibitemShut {NoStop}\bibitem [{\citenamefont {Shi}\ \emph {et~al.}(2026)\citenamefont {Shi},
  \citenamefont {Cano},\ and\ \citenamefont {Morales-Dur{\'a}n}}]{Shi2026}\BibitemOpen
  \bibfield  {author} {\bibinfo {author} {\bibfnamefont {J.}~\bibnamefont
  {Shi}}, \bibinfo {author} {\bibfnamefont {J.}~\bibnamefont {Cano}},\ and\
  \bibinfo {author} {\bibfnamefont {N.}~\bibnamefont {Morales-Dur{\'a}n}},\
  }\bibfield  {title} {\bibinfo {title} {Effects of {Berry} curvature on ideal
  fractional {Chern} insulator many-body gaps},\ }\href
  {https://doi.org/10.1103/sfm3-f1kp} {\bibfield  {journal} {\bibinfo
  {journal} {Phys. Rev. Research}\ }\textbf {\bibinfo {volume} {8}},\ \bibinfo
  {pages} {L022045} (\bibinfo {year} {2026})}\BibitemShut {NoStop}\bibitem [{\citenamefont {Moitra}\ and\ \citenamefont
  {Sodemann~Villadiego}(2026)}]{Moitra2026}\BibitemOpen
  \bibfield  {author} {\bibinfo {author} {\bibfnamefont {S.}~\bibnamefont
  {Moitra}}\ and\ \bibinfo {author} {\bibfnamefont {I.}~\bibnamefont
  {Sodemann~Villadiego}},\ }\bibfield  {title} {\bibinfo {title} {Instability
  of {Laughlin} fractional quantum {Hall} liquids into gapless power-law
  correlated states with continuous exponents in ideal {Chern} bands: Rigorous
  results from plasma mapping},\ }\href {https://doi.org/10.1103/3y87-96dw}
  {\bibfield  {journal} {\bibinfo  {journal} {Phys. Rev. B}\ }\textbf {\bibinfo
  {volume} {113}},\ \bibinfo {pages} {L161102} (\bibinfo {year}
  {2026})}\BibitemShut {NoStop}\bibitem [{\citenamefont {{OpenAI}}(2026)}]{OpenAI2026}\BibitemOpen
  \bibfield  {author} {\bibinfo {author} {\bibnamefont {{OpenAI}}},\ }\href
  {https://github.com/openai/math/tree/main/preprints/A-Fock-space-inequality-and-the-Laughlin-spectral-gap-September-24-2026}
  {\bibinfo {title} {A {Fock}-space inequality and the {Laughlin} spectral
  gap}},\ \bibinfo {howpublished} {OpenAI Math Release} (\bibinfo {year}
  {2026}),\ \bibinfo {note} {dated September 24; released October 6,
  2026}\BibitemShut {NoStop}\bibitem [{\citenamefont {Scheer}\ and\ \citenamefont
  {Khalaf}(2026)}]{Scheer2026}\BibitemOpen
  \bibfield  {author} {\bibinfo {author} {\bibfnamefont {M.~G.}\ \bibnamefont
  {Scheer}}\ and\ \bibinfo {author} {\bibfnamefont {E.}~\bibnamefont
  {Khalaf}},\ }\href {https://arxiv.org/abs/2610.18854v2} {\bibinfo {title} {A
  hierarchy of exact lower bounds on the {Laughlin} spectral gap}} (\bibinfo
  {year} {2026}),\ \bibinfo {note} {version 2},\ \Eprint
  {https://arxiv.org/abs/2610.18854} {arXiv:2610.18854 [cond-mat.str-el]}
  \BibitemShut {NoStop}\bibitem [{\citenamefont {Gao}(2026)}]{Gao2026}\BibitemOpen
  \bibfield  {author} {\bibinfo {author} {\bibfnamefont {Q.}~\bibnamefont
  {Gao}},\ }\href {https://arxiv.org/abs/2610.12700v1} {\bibinfo {title}
  {Improving the lower bound on the {Laughlin} gap: An {SDP} search}} (\bibinfo
  {year} {2026}),\ \bibinfo {note} {version 1},\ \Eprint
  {https://arxiv.org/abs/2610.12700} {arXiv:2610.12700 [cond-mat.str-el]}
  \BibitemShut {NoStop}\bibitem [{\citenamefont {Wang}\ \emph
  {et~al.}(2021{\natexlab{b}})\citenamefont {Wang}, \citenamefont {Zheng},
  \citenamefont {Millis},\ and\ \citenamefont {Cano}}]{WangZheng2021}\BibitemOpen
  \bibfield  {author} {\bibinfo {author} {\bibfnamefont {J.}~\bibnamefont
  {Wang}}, \bibinfo {author} {\bibfnamefont {Y.}~\bibnamefont {Zheng}},
  \bibinfo {author} {\bibfnamefont {A.~J.}\ \bibnamefont {Millis}},\ and\
  \bibinfo {author} {\bibfnamefont {J.}~\bibnamefont {Cano}},\ }\bibfield
  {title} {\bibinfo {title} {Chiral approximation to twisted bilayer graphene:
  Exact intravalley inversion symmetry, nodal structure, and implications for
  higher magic angles},\ }\href
  {https://doi.org/10.1103/PhysRevResearch.3.023155} {\bibfield  {journal}
  {\bibinfo  {journal} {Phys. Rev. Research}\ }\textbf {\bibinfo {volume}
  {3}},\ \bibinfo {pages} {023155} (\bibinfo {year}
  {2021}{\natexlab{b}})}\BibitemShut {NoStop}\bibitem [{\citenamefont {Haldane}(2011)}]{Haldane2011}\BibitemOpen
  \bibfield  {author} {\bibinfo {author} {\bibfnamefont {F.~D.~M.}\
  \bibnamefont {Haldane}},\ }\bibfield  {title} {\bibinfo {title} {Geometrical
  description of the fractional quantum {Hall} effect},\ }\href
  {https://doi.org/10.1103/PhysRevLett.107.116801} {\bibfield  {journal}
  {\bibinfo  {journal} {Phys. Rev. Lett.}\ }\textbf {\bibinfo {volume} {107}},\
  \bibinfo {pages} {116801} (\bibinfo {year} {2011})}\BibitemShut {NoStop}\bibitem [{\citenamefont {Qiu}\ \emph {et~al.}(2012)\citenamefont {Qiu},
  \citenamefont {Haldane}, \citenamefont {Wan}, \citenamefont {Yang},\ and\
  \citenamefont {Yi}}]{Qiu2012}\BibitemOpen
  \bibfield  {author} {\bibinfo {author} {\bibfnamefont {R.-Z.}\ \bibnamefont
  {Qiu}}, \bibinfo {author} {\bibfnamefont {F.~D.~M.}\ \bibnamefont {Haldane}},
  \bibinfo {author} {\bibfnamefont {X.}~\bibnamefont {Wan}}, \bibinfo {author}
  {\bibfnamefont {K.}~\bibnamefont {Yang}},\ and\ \bibinfo {author}
  {\bibfnamefont {S.}~\bibnamefont {Yi}},\ }\bibfield  {title} {\bibinfo
  {title} {Model anisotropic quantum {Hall} states},\ }\href
  {https://doi.org/10.1103/PhysRevB.85.115308} {\bibfield  {journal} {\bibinfo
  {journal} {Phys. Rev. B}\ }\textbf {\bibinfo {volume} {85}},\ \bibinfo
  {pages} {115308} (\bibinfo {year} {2012})}\BibitemShut {NoStop}\bibitem [{\citenamefont {Krishna}\ \emph {et~al.}(2019)\citenamefont
  {Krishna}, \citenamefont {Chen}, \citenamefont {Ippoliti},\ and\
  \citenamefont {Bhatt}}]{Krishna2019}\BibitemOpen
  \bibfield  {author} {\bibinfo {author} {\bibfnamefont {A.}~\bibnamefont
  {Krishna}}, \bibinfo {author} {\bibfnamefont {F.}~\bibnamefont {Chen}},
  \bibinfo {author} {\bibfnamefont {M.}~\bibnamefont {Ippoliti}},\ and\
  \bibinfo {author} {\bibfnamefont {R.~N.}\ \bibnamefont {Bhatt}},\ }\bibfield
  {title} {\bibinfo {title} {Interaction-dependent anisotropy of fractional
  quantum {Hall} states},\ }\href {https://doi.org/10.1103/PhysRevB.100.085129}
  {\bibfield  {journal} {\bibinfo  {journal} {Phys. Rev. B}\ }\textbf {\bibinfo
  {volume} {100}},\ \bibinfo {pages} {085129} (\bibinfo {year}
  {2019})}\BibitemShut {NoStop}\bibitem [{\citenamefont {Wang}\ \emph {et~al.}(2012)\citenamefont {Wang},
  \citenamefont {Narayanan}, \citenamefont {Wan},\ and\ \citenamefont
  {Zhang}}]{WangAnisotropy2012}\BibitemOpen
  \bibfield  {author} {\bibinfo {author} {\bibfnamefont {H.}~\bibnamefont
  {Wang}}, \bibinfo {author} {\bibfnamefont {R.}~\bibnamefont {Narayanan}},
  \bibinfo {author} {\bibfnamefont {X.}~\bibnamefont {Wan}},\ and\ \bibinfo
  {author} {\bibfnamefont {F.}~\bibnamefont {Zhang}},\ }\bibfield  {title}
  {\bibinfo {title} {Fractional quantum {Hall} states in two-dimensional
  electron systems with anisotropic interactions},\ }\href
  {https://doi.org/10.1103/PhysRevB.86.035122} {\bibfield  {journal} {\bibinfo
  {journal} {Phys. Rev. B}\ }\textbf {\bibinfo {volume} {86}},\ \bibinfo
  {pages} {035122} (\bibinfo {year} {2012})}\BibitemShut {NoStop}\end{thebibliography}

\begin{thebibliography}{11}\makeatletter
\providecommand \@ifxundefined [1]{\@ifx{#1\undefined}
}\providecommand \@ifnum [1]{\ifnum #1\expandafter \@firstoftwo
 \else \expandafter \@secondoftwo
 \fi
}\providecommand \@ifx [1]{\ifx #1\expandafter \@firstoftwo
 \else \expandafter \@secondoftwo
 \fi
}\providecommand \natexlab [1]{#1}\providecommand \enquote  [1]{``#1''}\providecommand \bibnamefont  [1]{#1}\providecommand \bibfnamefont [1]{#1}\providecommand \citenamefont [1]{#1}\providecommand \href@noop [0]{\@secondoftwo}\providecommand \href [0]{\begingroup \@sanitize@url \@href}\providecommand \@href[1]{\@@startlink{#1}\@@href}\providecommand \@@href[1]{\endgroup#1\@@endlink}\providecommand \@sanitize@url [0]{\catcode `\\12\catcode `\$12\catcode
  `\&12\catcode `\#12\catcode `\^12\catcode `\_12\catcode `\%12\relax}\providecommand \@@startlink[1]{}\providecommand \@@endlink[0]{}\providecommand \url  [0]{\begingroup\@sanitize@url \@url }\providecommand \@url [1]{\endgroup\@href {#1}{\urlprefix }}\providecommand \urlprefix  [0]{URL }\providecommand \Eprint [0]{\href }\providecommand \doibase [0]{https://doi.org/}\providecommand \selectlanguage [0]{\@gobble}\providecommand \bibinfo  [0]{\@secondoftwo}\providecommand \bibfield  [0]{\@secondoftwo}\providecommand \translation [1]{[#1]}\providecommand \BibitemOpen [0]{}\providecommand \bibitemStop [0]{}\providecommand \bibitemNoStop [0]{.\EOS\space}\providecommand \EOS [0]{\spacefactor3000\relax}\providecommand \BibitemShut  [1]{\csname bibitem#1\endcsname}\let\auto@bib@innerbib\@empty
\bibitem [{\citenamefont {Gao}(2026)}]{supp:Gao2026}\BibitemOpen
  \bibfield  {author} {\bibinfo {author} {\bibfnamefont {Q.}~\bibnamefont
  {Gao}},\ }\href {https://arxiv.org/abs/2610.12700v1} {\bibinfo {title}
  {Improving the lower bound on the {Laughlin} gap: An {SDP} search}} (\bibinfo
  {year} {2026}),\ \bibinfo {note} {version 1},\ \Eprint
  {https://arxiv.org/abs/2610.12700} {arXiv:2610.12700 [cond-mat.str-el]}
  \BibitemShut {NoStop}\bibitem [{\citenamefont {{OpenAI}}(2026)}]{supp:OpenAI2026}\BibitemOpen
  \bibfield  {author} {\bibinfo {author} {\bibnamefont {{OpenAI}}},\ }\href
  {https://github.com/openai/math/tree/main/preprints/A-Fock-space-inequality-and-the-Laughlin-spectral-gap-September-24-2026}
  {\bibinfo {title} {A {Fock}-space inequality and the {Laughlin} spectral
  gap}},\ \bibinfo {howpublished} {OpenAI Math Release} (\bibinfo {year}
  {2026}),\ \bibinfo {note} {dated September 24; released October 6,
  2026}\BibitemShut {NoStop}\bibitem [{\citenamefont {Scheer}\ and\ \citenamefont
  {Khalaf}(2026)}]{supp:Scheer2026}\BibitemOpen
  \bibfield  {author} {\bibinfo {author} {\bibfnamefont {M.~G.}\ \bibnamefont
  {Scheer}}\ and\ \bibinfo {author} {\bibfnamefont {E.}~\bibnamefont
  {Khalaf}},\ }\href {https://arxiv.org/abs/2610.18854v2} {\bibinfo {title} {A
  hierarchy of exact lower bounds on the {Laughlin} spectral gap}} (\bibinfo
  {year} {2026}),\ \bibinfo {note} {version 2},\ \Eprint
  {https://arxiv.org/abs/2610.18854} {arXiv:2610.18854 [cond-mat.str-el]}
  \BibitemShut {NoStop}\bibitem [{\citenamefont {Wang}\ \emph {et~al.}(2021)\citenamefont {Wang},
  \citenamefont {Zheng}, \citenamefont {Millis},\ and\ \citenamefont
  {Cano}}]{supp:WangZheng2021}\BibitemOpen
  \bibfield  {author} {\bibinfo {author} {\bibfnamefont {J.}~\bibnamefont
  {Wang}}, \bibinfo {author} {\bibfnamefont {Y.}~\bibnamefont {Zheng}},
  \bibinfo {author} {\bibfnamefont {A.~J.}\ \bibnamefont {Millis}},\ and\
  \bibinfo {author} {\bibfnamefont {J.}~\bibnamefont {Cano}},\ }\bibfield
  {title} {\bibinfo {title} {Chiral approximation to twisted bilayer graphene:
  Exact intravalley inversion symmetry, nodal structure, and implications for
  higher magic angles},\ }\href
  {https://doi.org/10.1103/PhysRevResearch.3.023155} {\bibfield  {journal}
  {\bibinfo  {journal} {Phys. Rev. Research}\ }\textbf {\bibinfo {volume}
  {3}},\ \bibinfo {pages} {023155} (\bibinfo {year} {2021})}\BibitemShut
  {NoStop}\bibitem [{\citenamefont {Vozmediano}\ \emph {et~al.}(2010)\citenamefont
  {Vozmediano}, \citenamefont {Katsnelson},\ and\ \citenamefont
  {Guinea}}]{supp:Vozmediano2010}\BibitemOpen
  \bibfield  {author} {\bibinfo {author} {\bibfnamefont {M.~A.~H.}\
  \bibnamefont {Vozmediano}}, \bibinfo {author} {\bibfnamefont {M.~I.}\
  \bibnamefont {Katsnelson}},\ and\ \bibinfo {author} {\bibfnamefont
  {F.}~\bibnamefont {Guinea}},\ }\bibfield  {title} {\bibinfo {title} {Gauge
  fields in graphene},\ }\href {https://doi.org/10.1016/j.physrep.2010.07.003}
  {\bibfield  {journal} {\bibinfo  {journal} {Phys. Rep.}\ }\textbf {\bibinfo
  {volume} {496}},\ \bibinfo {pages} {109} (\bibinfo {year}
  {2010})}\BibitemShut {NoStop}\bibitem [{\citenamefont {Gao}\ \emph {et~al.}(2023)\citenamefont {Gao},
  \citenamefont {Dong}, \citenamefont {Ledwith}, \citenamefont {Parker},\ and\
  \citenamefont {Khalaf}}]{supp:GaoStrain2023}\BibitemOpen
  \bibfield  {author} {\bibinfo {author} {\bibfnamefont {Q.}~\bibnamefont
  {Gao}}, \bibinfo {author} {\bibfnamefont {J.}~\bibnamefont {Dong}}, \bibinfo
  {author} {\bibfnamefont {P.}~\bibnamefont {Ledwith}}, \bibinfo {author}
  {\bibfnamefont {D.}~\bibnamefont {Parker}},\ and\ \bibinfo {author}
  {\bibfnamefont {E.}~\bibnamefont {Khalaf}},\ }\bibfield  {title} {\bibinfo
  {title} {Untwisting moir{\'e} physics: Almost ideal bands and fractional
  {Chern} insulators in periodically strained monolayer graphene},\ }\href
  {https://doi.org/10.1103/PhysRevLett.131.096401} {\bibfield  {journal}
  {\bibinfo  {journal} {Phys. Rev. Lett.}\ }\textbf {\bibinfo {volume} {131}},\
  \bibinfo {pages} {096401} (\bibinfo {year} {2023})}\BibitemShut {NoStop}\bibitem [{\citenamefont {de~Juan}\ \emph {et~al.}(2013)\citenamefont
  {de~Juan}, \citenamefont {Ma{\~n}es},\ and\ \citenamefont
  {Vozmediano}}]{supp:DeJuan2013}\BibitemOpen
  \bibfield  {author} {\bibinfo {author} {\bibfnamefont {F.}~\bibnamefont
  {de~Juan}}, \bibinfo {author} {\bibfnamefont {J.~L.}\ \bibnamefont
  {Ma{\~n}es}},\ and\ \bibinfo {author} {\bibfnamefont {M.~A.~H.}\ \bibnamefont
  {Vozmediano}},\ }\bibfield  {title} {\bibinfo {title} {Gauge fields from
  strain in graphene},\ }\href {https://doi.org/10.1103/PhysRevB.87.165131}
  {\bibfield  {journal} {\bibinfo  {journal} {Phys. Rev. B}\ }\textbf {\bibinfo
  {volume} {87}},\ \bibinfo {pages} {165131} (\bibinfo {year}
  {2013})}\BibitemShut {NoStop}\bibitem [{\citenamefont {Estienne}\ \emph {et~al.}(2023)\citenamefont
  {Estienne}, \citenamefont {Regnault},\ and\ \citenamefont
  {Cr{\'e}pel}}]{supp:Estienne2023}\BibitemOpen
  \bibfield  {author} {\bibinfo {author} {\bibfnamefont {B.}~\bibnamefont
  {Estienne}}, \bibinfo {author} {\bibfnamefont {N.}~\bibnamefont {Regnault}},\
  and\ \bibinfo {author} {\bibfnamefont {V.}~\bibnamefont {Cr{\'e}pel}},\
  }\bibfield  {title} {\bibinfo {title} {Ideal {Chern} bands are {Landau}
  levels in curved space},\ }\href
  {https://doi.org/10.1103/PhysRevResearch.5.L032048} {\bibfield  {journal}
  {\bibinfo  {journal} {Phys. Rev. Research}\ }\textbf {\bibinfo {volume}
  {5}},\ \bibinfo {pages} {L032048} (\bibinfo {year} {2023})}\BibitemShut
  {NoStop}\bibitem [{\citenamefont {Qiu}\ \emph {et~al.}(2012)\citenamefont {Qiu},
  \citenamefont {Haldane}, \citenamefont {Wan}, \citenamefont {Yang},\ and\
  \citenamefont {Yi}}]{supp:Qiu2012}\BibitemOpen
  \bibfield  {author} {\bibinfo {author} {\bibfnamefont {R.-Z.}\ \bibnamefont
  {Qiu}}, \bibinfo {author} {\bibfnamefont {F.~D.~M.}\ \bibnamefont {Haldane}},
  \bibinfo {author} {\bibfnamefont {X.}~\bibnamefont {Wan}}, \bibinfo {author}
  {\bibfnamefont {K.}~\bibnamefont {Yang}},\ and\ \bibinfo {author}
  {\bibfnamefont {S.}~\bibnamefont {Yi}},\ }\bibfield  {title} {\bibinfo
  {title} {Model anisotropic quantum {Hall} states},\ }\href
  {https://doi.org/10.1103/PhysRevB.85.115308} {\bibfield  {journal} {\bibinfo
  {journal} {Phys. Rev. B}\ }\textbf {\bibinfo {volume} {85}},\ \bibinfo
  {pages} {115308} (\bibinfo {year} {2012})}\BibitemShut {NoStop}\bibitem [{\citenamefont {Krishna}\ \emph {et~al.}(2019)\citenamefont
  {Krishna}, \citenamefont {Chen}, \citenamefont {Ippoliti},\ and\
  \citenamefont {Bhatt}}]{supp:Krishna2019}\BibitemOpen
  \bibfield  {author} {\bibinfo {author} {\bibfnamefont {A.}~\bibnamefont
  {Krishna}}, \bibinfo {author} {\bibfnamefont {F.}~\bibnamefont {Chen}},
  \bibinfo {author} {\bibfnamefont {M.}~\bibnamefont {Ippoliti}},\ and\
  \bibinfo {author} {\bibfnamefont {R.~N.}\ \bibnamefont {Bhatt}},\ }\bibfield
  {title} {\bibinfo {title} {Interaction-dependent anisotropy of fractional
  quantum {Hall} states},\ }\href {https://doi.org/10.1103/PhysRevB.100.085129}
  {\bibfield  {journal} {\bibinfo  {journal} {Phys. Rev. B}\ }\textbf {\bibinfo
  {volume} {100}},\ \bibinfo {pages} {085129} (\bibinfo {year}
  {2019})}\BibitemShut {NoStop}\bibitem [{\citenamefont {Wang}\ \emph {et~al.}(2012)\citenamefont {Wang},
  \citenamefont {Narayanan}, \citenamefont {Wan},\ and\ \citenamefont
  {Zhang}}]{supp:WangAnisotropy2012}\BibitemOpen
  \bibfield  {author} {\bibinfo {author} {\bibfnamefont {H.}~\bibnamefont
  {Wang}}, \bibinfo {author} {\bibfnamefont {R.}~\bibnamefont {Narayanan}},
  \bibinfo {author} {\bibfnamefont {X.}~\bibnamefont {Wan}},\ and\ \bibinfo
  {author} {\bibfnamefont {F.}~\bibnamefont {Zhang}},\ }\bibfield  {title}
  {\bibinfo {title} {Fractional quantum {Hall} states in two-dimensional
  electron systems with anisotropic interactions},\ }\href
  {https://doi.org/10.1103/PhysRevB.86.035122} {\bibfield  {journal} {\bibinfo
  {journal} {Phys. Rev. B}\ }\textbf {\bibinfo {volume} {86}},\ \bibinfo
  {pages} {035122} (\bibinfo {year} {2012})}\BibitemShut {NoStop}\end{thebibliography}
